\documentclass[12pt]{article}
\usepackage{amsmath,amssymb,bm,epsfig,color,graphicx}
\renewcommand{\thefootnote}{\fnsymbol{footnote}}

\newcommand{\hka}{\hat{\kappa}}

\newcommand{\ham}{\hat{a}_m}
\newcommand{\hah}{\hat{a}_H}
\newcommand{\hbh}{\hat{b}_H}
\newcommand{\hSo}{\hat{S}_1}
\newcommand{\hSt}{\hat{S}_2}
\newcommand{\hSz}{\hat{S}_0}
\newcommand{\hph}{\hat{\phi}}
\newcommand{\dI}{\delta I}

\begin{document}

\title{
\begin{flushright}
\begin{minipage}{0.2\linewidth}
\normalsize
WU-HEP-18-11\\*[50pt]
\end{minipage}
\end{flushright}
{\Large \bf 
Affleck-Dine baryogenesis\\ in the SUSY Dine-Fischler-Srednicki-Zhitnitsky\\ axion model without R-parity
\\*[20pt]}}

\author{
Kensuke Akita$^{1}$\footnote{
E-mail address: kensuke@th.phys.titech.ac.jp}\ and\ 
Hajime Otsuka$^{2}$\footnote{
E-mail address: h.otsuka@aoni.waseda.jp}\\*[20pt]
$^1${\it \normalsize
Department of Physics, Tokyo Institute of Technology,
Tokyo 152-8551, Japan} \\*[5pt]
$^2${\it \normalsize 
Department of Physics, Waseda University, 
Tokyo 169-8555, Japan} \\*[50pt]
}

\date{
\centerline{\small \bf Abstract}
\begin{minipage}{0.9\linewidth}
\medskip 
\medskip 
\small
\ \ \ \ We investigate the baryon asymmetry in the supersymmetry Dine-Fischler-Srednicki-Zhitnitsky axion model without R-parity. 
It turns out that the R-parity violating terms economically explain the atmospheric mass-squared difference of neutrinos 
and the appropriate amount of baryon asymmetry through the Affleck-Dine mechanism. 
In this model, the axion is a promising candidate for the dark matter and the axion isocurvature perturbation is suppressed due to the large field values of Peccei-Quinn fields.  
Remarkably, in some parameter regions 
explaining the baryon asymmetry and the axion dark matter abundance, the proton decay will be explored in future experiments.
\end{minipage}
}

\begin{titlepage}
\maketitle
\thispagestyle{empty}
\clearpage
\tableofcontents
\end{titlepage}

\renewcommand{\thefootnote}{\arabic{footnote}}
\setcounter{footnote}{0}

\section{Introduction}

The Standard Model (SM) of elementary particle physics is very consistent with the experimental data up to TeV scale. 
However, neutrino oscillations reported in the SuperKamiokande~\cite{NO}, strong CP problem, the excess of baryon over antibaryon in the current Universe, and the absence of a dark matter candidate in the SM  indicate new physics beyond the SM. 
To explain these phenomenological problems, we focus on the supersymmetry (SUSY) as an extension of the SM.

So far, several models have been proposed to explain the nonvanishing neutrino masses as represented by the seesaw mechanism, introducing the heavy right-handed Majorana neutrinos in the SM~\cite{Seesaw1,Seesaw2,Seesaw3,Seesaw4,Seesaw5}. 
Among supersymmetric models, the R-parity violating SUSY scenario is the simplest approach to 
explain the tiny neutrino masses~\cite{RPVNM}.\footnote{For more details, see, e.g., Ref.~\cite{Barbier:2004ez}.} 
Because of the R-parity violation, the lepton number violating interactions generate the neutrino masses without introducing a new particle in the framework of SUSY. 
In addition, the R-parity violating interactions help us to avoid cosmological gravitino and moduli problems such as the 
overabundance of the lightest supersymmetric particle generated by the gravitino and moduli decays~\cite{Endo:2006zj,Nakamura:2006uc,Dine:2006ii}, although the moduli are required to be heavier than ${\cal O}(100)$\ TeV to realize a successful big-bang nucleosynthesis. 
However, the sizable R-parity violating interactions cause several cosmological and phenomenological problems such as 
the unobservable proton decay, undetectable collider signatures of supersymmetric particles and washing out the primordial baryon asymmetry. 
Thus it is necessary to explain the smallness of R-parity violating interactions.

To explain the smallness of R-parity violating interactions, we focus on the SUSY 
Dine-Fischler-Srednicki-Zhitnitsky (DFSZ) axion model~\cite{DFSZaxion}. 
In this model, the strong CP problem in the quantum chromodynamics (QCD) can be solved using the Peccei-Quinn (PQ) mechanism~\cite{PQ}. 
The Nambu-Goldstone boson called axion~\cite{axion1, axion2} appears through 
the spontaneous symmetry breaking of global $U(1)_{\rm PQ}$ and dynamically cancels the CP phases in QCD. 
Furthermore this axion is a promising candidate for dark matter and its coherent oscillation explains the current dark matter 
abundance~\cite{axionDM}. 
Note that the SUSY DFSZ axion model controls the size of the $\mu$-term and the R-parity violating couplings by $U(1)_{{\rm PQ}}$ symmetry. 
It is then important to explore whether or not enough baryon asymmetry is realized in such an extension of the SM. 
In an inflationary era, the primordial baryon asymmetry is diluted away by the accelerated expansion of the Universe~\cite{inflation1,inflation2}. 
The baryon asymmetry, in particular the baryon to photon ratio $\eta\simeq 5\times10^{-10}$ required by the big-bang nucleosynthesis and cosmic microwave background~\cite{BBNCMB}, should be created after inflation. 
To obtain the sizable baryon asymmetry after inflation, we study the Affleck-Dine (AD) mechanism~\cite{Affleck:1984fy,Dine:1995kz} 
in the SUSY DFSZ axion model without R-parity. 
Since PQ fields couple to the baryon/lepton number violating terms to control the size of the R-parity violating couplings in this setup, it is nontrivial that the AD mechanism produces the appropriate amount of baryon asymmetry.

The aim of this work is to reveal the origin of baryon asymmetry in the SUSY DFSZ axion model 
without R-parity in comparison with the usual supersymmetric and R-parity conserving axion models. 
In contrast to the R-parity violating AD mechanism proposed in Ref.~\cite{ADwithR-parity},
the R-parity violating terms in our model couple to PQ fields. The dynamics of PQ fields affects the magnitude of R-parity violation and the amount of baryon asymmetry.\footnote{For the AD leptogenesis scenario with a varying PQ scale, see Ref.~\cite{Bae:2016zym}. Also see Ref.~\cite{Akita:2017ecc}, where the axion inflaton affects the dynamics of AD field.} 
In this regard, our model is different from Ref.~\cite{ADwithR-parity}.
We find that there exist some parameter regions, explaining the current baryon asymmetry, 
the axion dark matter abundance, and the smallness of $\mu$- and R-parity violating terms. 
It is also consistent with the unobservable proton decay, undetectable collider signatures of supersymmetric particles, washout of the baryon asymmetry
and one of the two mass-squared differences of neutrinos, corresponding to the atmospheric mass-squared difference of neutrinos.\footnote{We leave the derivation of another mass-squared difference called the solar mass-squared difference of neutrinos to one of our future works.}
Remarkably, in this model, the axion isocurvature perturbation is suppressed because of the enhancement of the PQ breaking scale 
in the early Universe. 
Furthermore, in some parameter regions 
explaining baryon asymmetry and axion dark matter abundance, the proton decay will be explored in future experiments such as the HyperKamiokande~\cite{HyperK}.

In particular, this model is economical in the point of view of explaining the atmospheric neutrino mass data, the baryon asymmetry and the strong CP problem in supersymmetric models. This is because we do not introduce a new field to explain the atmospheric neutrino mass data and the baryon asymmetry in the SUSY DFSZ axion model and we do not impose the discrete symmetry called R-parity which is usually imposed in supersymmetric models.

This paper is organized as follows. In Sec.~2, we consider the SUSY DFSZ axion model without R-parity and show constraints on 
 the R-parity violating couplings. 
In Sec.~3 we investigate the AD baryogenesis in more detail. Then we study the dynamics of the AD and PQ fields and estimate the baryon asymmetry, taking account of the dilution of saxion decay.  The axion isocurvature perturbation is then suppressed. 
Finally, we conclude in Sec.~4. 

\section{Model and constraints from experiments} 
\label{sec:2}

\subsection{Setup}

In this paper, we consider the SUSY DFSZ axion model in which 
the global $U(1)_{\rm PQ}$ symmetry and the PQ fields $S_0, S_1, S_2$ are introduced 
adding to the Minimal Supersymmetric SM (MSSM).~\footnote{See, as a review, Ref.~\cite{SUSY}.}
The gauge symmetry of the MSSM does not prohibit 
either the baryon or the lepton number violating coupling in the superpotential. 
So-called R-parity is often assumed in the MSSM to control these couplings
since the proton lifetime easily becomes shorter than the observational bounds. 
In the SUSY DFSZ model, the $U(1)_{\rm PQ}$ symmetry plays this role instead 
of the conventional R-parity. 
It will be shown that the charge assignment of the $U(1)_{\rm PQ}$ 
explains not only the smallness of the baryon and lepton number violating couplings
but also the size of the $\mu$-parameter.

We assume the following superpotential:
\begin{align}
W&=W_{\rm {MSSM}}+W_{R_p\hspace{-3.5mm}/}+W_{\rm {PQ}}, \\
W_{\rm {MSSM}}&= y^u_{ij}\overline{u}_iQ_jH_u-y^d_{ij}\overline{d}_iQ_jH_d-y^e_{ij}\overline{e}_iL_jH_d+\frac{y_0 S_1^2}{M_P}H_uH_d, \label{eq:WMSSM}\\
W_{R_p\hspace{-3.5mm}/}&= \frac{y_i S_1^3}{M_P^2}L_iH_u+\frac{\gamma_{ijk} S_1}{M_P}L_iL_j\overline{e}_k+\frac{\gamma_{ijk}' S_1}{M_P}L_iQ_j
\overline{d}_k+\frac{\gamma_{ijk}'' S_1^3}{M_P^3}\overline{u}_i\overline{d}_j\overline{d}_k, \label{SRPV}\\ 
W_{\rm {PQ}}&= \kappa S_0(S_1S_2-f^2),
\end{align}
where $i,j,k=1,2,3$ are family indices. 
The Higgs fields, quarks, and leptons are the usual representations 
of the MSSM gauge group,  
and the PQ fields are singlet under the MSSM gauge group. 
$y_0, y_i, \gamma_{ijk}, \gamma'_{ijk}, \gamma''_{ijk}, \kappa$ 
are dimensionless parameters which are typically of $\mathcal{O}(1)$.

\begin{table}[tb]
  \centering
  \begin{tabular}{|r|r|r|r|r|r|r|r|r|r|r|r|} \hline
&$S_0$&$S_1$ & $S_2$ & $H_u$ & $H_d$ & $\overline{u}_i$ & $\overline{d}_i$ 
& $Q_i$ & $\overline{e}_i$ & $L_i$  \\  \hline
    $PQ$ & $0$ & $1$ & $-1$ & $-1$ & $-1$ & $-1$ & $-1$ & $2$ & $3$ & $-2$ \\  \hline
    $B$  & $0$ & $0$ & $0$ & $0$ & $0$ & $-1/3$ & $-1/3$ & $1/3$ & $0$ & $0$\\  \hline
  \end{tabular}
  \caption{PQ charges and baryon numbers of fields.}
  \label{tab-PQcharge}
\end{table}

The PQ charges and also the baryon numbers of each field are listed 
in Table~\ref{tab-PQcharge}.
$f$ in $W_{\rm PQ}$ is the present PQ breaking scale 
and $M_P \simeq 2.4 \times 10^{18}~{\rm GeV}$ is the reduced Planck mass. 
When we arrange the PQ fields to $S_1=Se^{A/f}, S_2=Se^{-A/f}$, 
$A$ behaves as the axion superfield.  
$W_{\rm PQ}$ induces a spontaneous symmetry breaking of $U(1)_{\rm PQ}$ at the minimum $\langle S \rangle \simeq f$. 
The present effective superpotential after the symmetry breaking 
is given by 
\begin{align}
W_{\rm eff}=& W_{\rm Yukawa} + \mu_0e^{2A/f}H_uH_d+\mu_ie^{3A/f}L_iH_u \nonumber \\
&+\lambda_{ijk}e^{A/f}L_iL_j\overline{e}_k+\lambda'_{ijk}e^{A/f}L_iQ_j\overline{d}_k+\lambda''_{ijk}e^{3A/f}\overline{u}_i\overline{d}_j\overline{d}_k,
\end{align}
where $W_{\rm Yukawa}$ denotes the Yukawa coupling terms in Eq.~(\ref{eq:WMSSM}) and 
\begin{align}
\mu_0&=\frac{y_0f^2}{M_P},\  \mu_i=\frac{y_i f^3}{M_P^2}, \nonumber \\
 \lambda_{ijk}&=\gamma_{ijk}\left(\frac{f}{M_P}\right),\ 
 \lambda'_{ijk}=\gamma'_{ijk}\left(\frac{f}{M_P}\right),\ 
\lambda''_{ijk}=\gamma''_{ijk} \left(\frac{f}{M_P} \right)^3
\end{align}
are defined. 
For$\ f=9.0 \times 10^{11}~{\rm GeV}, y_0= y_i =1$, $\gamma_{ijk} = \gamma'_{ijk} =1 $ and $ \gamma''_{ijk} = 1$, we obtain
\begin{align}
\mu_0 &=3.4\times10^5~{\rm GeV},\ \mu_i=1.3 \times 10^{-1}~{\rm GeV},\nonumber \\ 
\lambda_{ijk}&=3.8\times 10^{-7},\  \lambda_{ijk}' =3.8 \times10^{-7},\  \lambda_{ijk}'' =5.3 \times10^{-20}.
\label{RPV}
\end{align}
Thus, the smallness of the $\mu$-parameter is explained 
by the Kim-Nilles mechanism~\cite{Kim-Nilles},  
and at the same time the R-parity violating couplings are suppressed by the Froggatt-Nielsen mechanism~\cite{Froggatt:1978nt} 
by the $U(1)_{\rm PQ}$ assignment.
The baryon number violating coupling constants $\lambda''_{ijk}$ are highly suppressed, whereas 
the lepton number is sizably violated. 
These facts lead to stability under the single nucleon decay~\cite{nucleondecay} and 
suppress effects from the washing out of the baryon asymmetry~\cite{Bwashout1,Bwashout2,Bwashout3,Bwashout4}.
It may be possible that one imposes flavor-dependent assignment of the $U(1)_{\rm PQ}$, which is related to the flaxion models proposed 
by Refs.~\cite{Flaxion1,Flaxion2,Flaxion3}.

\subsection{Constraints on the R-parity violating couplings}

We briefly review experimental constraints on the R-parity violating terms.  In the following, we enumerate only
some severe bounds on the R-parity violating couplings. A more comprehensive review is written in Ref.~\cite{Barbier:2004ez}.
We comment on the relation between these constraints and our models in Sec.~2.2.3.

\subsubsection{Proton decay}

The observations of single nucleon decay give an important constraint 
on the trilinear R-parity violating terms~\cite{nucleondecay}. 
A coexistence of the baryon and the lepton number violating operators 
induces a decay $p\rightarrow \pi^0l^{+}$ mediated by a $\tilde{d}_R$ squark 
in an s-channel which has not been detected so far, thus giving the upper bounds on the trilinear R-parity violating couplings~\cite{nucleondecay2, nucleondecay3},
\begin{align}
|\lambda'_{imk}\lambda''^{\ast}_{11k}|<\mathcal{O}(1)\times 10^{-25}
\biggl(\frac{m_{\tilde{d}}}{5~{\rm TeV}}\biggl)^2,
\label{ND}
\end{align}
where $i,k=1,2,3,m=1,2$, and $m_{\tilde{d}}$ is the typical down-type squark mass. 
The upper bounds are very severe, 
but the $U(1)_{\rm PQ}$-charge assignment leads to $\lambda' \lambda'' \sim 10^{-26}$, 
so the proton lifetime is longer than the bound.  

\subsubsection{Baryon washout}

Other upper bounds on the trilinear R-parity violating terms come 
from the observation of baryon asymmetry. If the baryon asymmetry is produced before the electroweak phase transition and the sphaleron process~\cite{Sphaleron1} is in the thermal equilibrium, the R-parity violating terms would erase the existing baryon asymmetry~\cite{Bwashout1,Bwashout2,Bwashout3,Bwashout4}. 
To avoid these effects, the trilinear R-parity violating couplings are upper bounded by
\begin{align}
\lambda , \lambda', \lambda'' < 4\times 10^{-7}\biggl(\frac{m_{\tilde{f}}}{1~{\rm TeV}} \biggl)^{1/2},
\label{BWO}
\end{align}
where $m_{\tilde{f}}$ is the typical mass of the sfermions.

\subsubsection{Neutrino masses}

We next focus on the bilinear R-parity violating couplings $\mu_i$.  Constraints on the R-parity violating couplings come from the cosmological observations on neutrino masses because the bilinear R-parity violation gives rise to the neutrino mass.  
When we choose the basis in which the vacuum expectation values (VEVs) of sneutrinos vanish and the Yukawa couplings of charged lepton are diagonal, 
the effective neutrino mass matrix at tree level is generated by the bilinear R-parity violating terms~\cite{NMRPV},
\begin{align}
M^{\nu}_{\rm tree}\simeq -\frac{m_{\nu {\rm tree}}}{\Sigma^3_{i=1} \mu^2_i}
\left(
\begin{array}{ccc}
\mu^2_1 & \mu_1\mu_2 & \mu_1\mu_3 \\
\mu_1\mu_2 & \mu^2_2 & \mu_2\mu_3 \\
\mu_1\mu_3 & \mu_3\mu_2 & \mu^2_3 \\
\end{array}
\right).
\end{align}
The size of neutrino mass $m_{\nu_{\rm tree}}$ is given by 
\begin{align}
m_{\nu_{\rm tree}}\simeq \frac{m_Z^2 (\cos^2 \theta_W M_1+\sin^2 \theta_WM_2)}{M_1M_2(1+\tan^2\beta)}\tan^2\xi,
\end{align}
where $m_Z,\ M_1$, $\ M_2$ are the $Z$ boson mass, the bino masse and the wino mass respectively, and $\theta_W$ is the Weinberg angle. $\tan\beta=v_u/v_d$ is the ratio between the VEVs of two Higgs fields  $v_u,\ v_d$ and
\begin{align}
\tan ^2\xi \simeq \frac{\mu^2_1+\mu^2_2+\mu^2_3}{\mu_0^2}.
\end{align}
Since the rank of the mass matrix is 1, one of the neutrinos acquires a mass at tree level.  Although the other two neutrinos acquire masses by quantum corrections~\cite{Takayama:1999pc,Hirsch:2000ef}, the detailed analysis of the other neutrino masses is beyond the scope of this paper. 
Indeed, one of the neutrino mass-squared differences can be explained by the bilinear R-parity violating terms, observed by the atmospheric neutrino observation~\cite{NMmass}
\begin{align}
\sqrt{\Delta m_A ^2} \simeq 5 \times 10^{-2}~{\rm eV}.
\end{align}
If we explain the atmospheric neutrino observation without tuning of the dimensionless parameters, the size of neutrino mass $m_{\nu_{\rm tree}}$ is constrained as $m_{\nu_{\rm tree}} \lesssim 5 \times 10^{-2}\ {\rm eV}$ which leads to the constraint on the bilinear R-parity violating couplings, 
\begin{align}
\sum_i \mu_i^2 \lesssim 6.0\times 10^{-12}\ {\rm eV}(1+\tan^2\beta)\biggl(\frac{M_2}{1\ {\rm TeV}} \biggl)\mu_0^2.
\label{eq:bilinearneutrino}
\end{align}

In addition, the cosmological bound on neutrino masses $\sum_i m_{\nu_i} \lesssim 0.11\ {\rm eV}$ $(m_{\nu_{\rm tree}} \lesssim 0.11\ {\rm eV})$~\cite{Planck2018} leads to another constraint on the bilinear R-parity violating couplings,
\begin{align}
\sum_i \mu_i^2 \lesssim 1.3\times 10^{-11}(1+\tan^2\beta)\biggl(\frac{M_2}{1\ {\rm TeV}}\biggl)\mu_0^2.
\label{eq:CMBneutrino}
\end{align}

In our model, the R-parity violating couplings (\ref{RPV}) avoid the constraints of Eqs.~(\ref{ND}), (\ref{BWO}), (\ref{eq:bilinearneutrino}), and (\ref{eq:CMBneutrino}). Furthermore, the bilinear R-parity violating couplings can explain atmospheric mass-squared difference of neutrinos in some region of dimensionless parameters.
Furthermore, in the low-scale SUSY scenario, we may observe the single nucleon decay for $f\simeq 10^{12}\ {\rm GeV}$ 
in future experiments such as HyperKamiokande~{\cite{HyperK}}.

In addition, values in Eqs.(\ref{RPV}) are derived under the choice of $f=9.0 \times10^{11}\ {\rm GeV}$ and hence is, in general, dependent on $f$. One can derive the bound on $f$ from constraints on the R-parity violating terms under $y, \gamma, \gamma', \gamma''=1$. 
If the masses of all of the supersymmetric particles are the same, the severest constraint on $f$ comes from Eq. (\ref{BWO}),
\begin{align}
f < 9.6 \times 10^{11}\ {\rm GeV}\biggl(\frac{m_{\tilde{f}}}{1\ {\rm TeV}} \biggl)^{1/2}.
\end{align}




\section{Affleck-Dine baryogenesis in the SUSY DFSZ axion model with the R-parity violating terms}
\label{sec:2}

In this section, 
we study the Affleck-Dine mechanism exploiting the R-parity violating terms  
in the SUSY DFSZ axion model. A notable point of our scenario is that the AD field couples to the PQ field 
and their dynamics are affected by each other.
This behavior makes a different amount of baryon asymmetry from the conventional one. In the following, we will take parameters which are consistent with the allowed region to explain the atmospheric neutrino mass-squared difference in Sec.~2.3.

\subsection{Affleck-Dine mechanism}
\label{sec:2_2}

Here and in the following, we consider the Affleck-Dine baryogenesis exploiting 
one of the $\overline{u}_i\overline{d}_j\overline{d}_k$ directions of the scalar potential.\footnote{If we impose the lepton number of $S_1$ as $\frac{1}{3}$, the AD baryogenesis via $\overline{u}_i\overline{d}_j\overline{d}_k$ cannot work. However, it may be possible that a certain baryon asymmetry is produced by other directions of the potential, e.g. $L_iQ_j\overline{d}_k$, based on a charge assignment different from the one in Table~\ref{tab-PQcharge} The detailed calculation is one of the future works.}. We consider one of the $\overline{u}\overline{d}\overline{d}$ D-flat directions as the so-called AD field namely 
\begin{align}
\overline{u}=\frac{1}{\sqrt{3}}
      \phi 
  ,\quad
\overline{d}=\frac{1}{\sqrt{3}}
        \phi.
\end{align}
In the next subsection, the potential for the AD field and the PQ fields is derived.  
The dynamics of  the AD/PQ fields during inflation is studied in Sec.~\ref{subsec:3_1_2}. 
After inflation, we show their dynamics in an era $H>m_\phi $ ($H\simeq m_\phi$) in Sec.~\ref{subsec:3_1_3} (\ref{subsec:3_1_4}). 
The resultant baryon asymmetry is extracted in Sec.~\ref{subsec:3_1_4}. 
We will apply the following method to the Affleck-Dine mechanism using other D-flat directions. However, it is especially nontrivial for the Affleck-Dine mechanism via the $LH_u$ directions based on the charge assignment in Table~\ref{tab-PQcharge}. This is because the $LH_u$ directions are related to the $\mu$-term($H_uH_d$ direction), which gives a non-negligible contribution to the scalar potential for the AD field in comparison to the conventional case. The potentials of other D-flat directions, including the $\bar{u}\bar{d}\bar{d}$ direction treated in this paper, do not receive this contribution.


\subsubsection{Potential for the AD/PQ fields}
\label{subsec:3_1_1}

We assume that the other D-flat directions have positive mass terms,  
so that we ignore effects from these directions in the following discussion. 
Note that some other D-flat directions coupled with the PQ fields 
will be available for the Affleck-Dine mechanism.  
For more details, see Appendix A, where we discuss 
the AD mechanism with general couplings 
between PQ field $S_1$ and AD field $\phi$.  
The generalized setup can be applied to the R-parity conserving case. 
The potential of the AD field and PQ fields depends on SUSY-breaking scenarios and 
it is important for the AD mechanism how these fields couple to the inflaton.  
In this paper, we assume the gravity-(or anomaly-)mediated SUSY-breaking scenario, together with the $F$-term inflation. 
In the gauge-mediated SUSY-breaking scenario, it is nontrivial that the AD mechanism works in this model. This is because the scalar potential is modified by the effect of the gauge-mediated supersymmetry breaking~\cite{ADgauge}.

Let us consider the following superpotential:
\begin{align}
W&=W_{\rm inf}(I)+W_{R_p\hspace{-3.5mm}/}\ (S_1,\phi)+W_{\rm {PQ}}
        + W_{\rm mix}, \nonumber \\
W_{R_p\hspace{-3.5mm}/}\ (S_1,\phi)&=-\frac{\gamma S_1^3\phi^3}{3M_P^3},\quad
W_{\rm {PQ}}= \kappa S_0(S_1S_2-f^2), 
\end{align}
where $W_{\rm inf}(I)$ is the superpotential for the inflaton $I$,  
$W_{R_p\hspace{-3.5mm}/}\ (\phi)$ originates from the last term in Eq.(\ref{SRPV}) 
, and 
$W_{\rm mix}$ stands for possible mixing terms between the inflaton and the AD/PQ fields.   
Here $\kappa$ is a coupling constant. Since the $\overline{u}\overline{d}\overline{d}$ direction has a flavor dependence, $\gamma$ represents for one of $\gamma''_{ijk}$. 

The supergravity scalar potential is given by~\cite{SUGRApotential}
\begin{align}
V=e^{K/M_P^2}\biggl[(D_aW)K^{a\overline{b}}(D_{\overline{b}} \overline{W})
                            -\frac{3}{M_P^2}|W|^2 \biggl],
\end{align}
with
\begin{align}
D_aW=\frac{\partial W}{\partial \Phi^a}+\frac{\partial K}{\partial \Phi^a}\frac{W}{M_P^2},
\quad
K^{a\overline{b}}=\biggl(\frac{\partial^2 K}{\partial \Phi^a \partial \Phi^{b{\ast}}}\biggl)^{-1}, 
\end{align}
where $\Phi^{a,b} = I, \phi, S_0, S_1, S_2$, and  
$K$ is the K$\ddot{{\rm a}}$hler potential. 
We assume the following K$\ddot{{\rm a}}$hler potential 
with nonminimal couplings, 
\begin{align}
K=
\sum_a \Phi^{a\dag} \Phi^a  
+\frac{\alpha}{M_P^2}\phi^{\dag}\phi I^{\dag}I+\frac{\beta}{M_P^2}S_1^{\dag}S_1I^{\dag}I+\mathcal{O}(M_P^{-3}),
\label{KP}
\end{align}
where the coupling constants $\alpha, \beta$ are introduced. 
Here  $S_0$ and $S_2$ have only the minimal K$\ddot{\rm a}$hler potential. 

The inflaton scalar potential is given by 
\begin{align}
V_{\rm inf} \simeq e^{K/M_P^2}\biggl( F_IK^{I\overline{I}}F^{\ast}_{\overline{I}}-\frac{3}{M_P^2}|W_{\rm inf}(I)|^2 \biggl).
\label{eq-Vinf}
\end{align}
If $\alpha,\beta \gtrsim 1$, the Hubble-induced mass terms are provided by the $F$-term of the inflaton,
\begin{align}
\label{eq-Vhub}
V_{\rm Hubble}\simeq 
    c_0H^2|S_0|^2-c_1H^2|S_1|^2+c_2H^2|S_2|^2-c_3H^2|\phi|^2,
\end{align}
where $c_0,c_1,c_2,c_3$ are positive constants and $H$ is the Hubble parameter.
The signs of the mass terms indicate 
that $S_0$ and $S_2$ acquire large positive masses during inflation. 
There are also soft SUSY-breaking mass terms, 
\begin{align}
\label{eq-Vsoft}
V_{\rm soft}=m_0^2|S_0|^2+m_1^2|S_1|^2+m_2^2|S_2|^2+m_{\phi}^2|\phi|^2.
\end{align}

Next, let us consider the contributions 
from $W_{R_p\hspace{-3.5mm}/}\ (S_1,\phi)$, $W_{PQ}$ and $W_{\rm mix}$. 
The $F$-term scalar potential is given by 
\begin{align}
V_{\rm F}=|\kappa|^2|S_1S_2-f^2|^2+\biggl|\kappa S_0S_2-\frac{\gamma S_1^2\phi^3}{M_P^3} \biggl|^2+|\kappa|^2|S_0S_1|^2+\frac{|\gamma|^2|S_1|^6|\phi|^4}{M_P^6}.
\label{eq-VF}
\end{align}
In our study, we assume that $|I| \ll M_P$ 
and the dynamics of the inflaton is basically separated 
from those of  the AD/PQ fields. 
Even in this case, the $F$-term of the inflaton $F_I$ 
may give significant effects to the dynamics of the AD/PQ fields. 
Let us consider the following mixing superpotential:
\begin{align}
\label{Vmix}
W_{\rm mix}=\alpha'\frac{I}{M_P}W_{R_p\hspace{-3.5mm}/}\ (S_1,\phi)
                    +\beta'\frac{I}{M_P}W_{\rm PQ}, 
\end{align}
where $\alpha',\beta'$ are coupling constants. 
This superpotential induces the potential, 
\begin{align}
\biggl(\alpha'W_{R_p\hspace{-3.5mm}/}\ (S_1,\phi)+\beta'W_{\rm PQ}\biggl)\frac{F_I^{\ast}}{M_P}+{\rm H.c.},
\end{align}
which are not suppressed even for $|I| \ll M_P$. 
Therefore we expect that there are Hubble-induced couplings in the scalar potential. 
In addition, there are soft SUSY-breaking terms, generating from the mixing superpotential, 
%
\begin{align}
V_{\rm A}=& (a_HH +a_mm_{3/2} ) W_{R_p\hspace{-3.5mm}/}\ (S_1,\phi)
                   +(b_HH + b_mm_{3/2} )W_{\rm PQ}+{\rm H.c.} \nonumber \\
=& -(a_HH +a_mm_{3/2} ) \frac{\gamma S_1^3\phi^3}{3M_P^3}
    +(b_HH + b_mm_{3/2} )\kappa S_0(S_1S_2-f^2)+ \text{H.c.},
\end{align}
where $a_H, b_H$ stand for the Hubble-induced couplings 
and $a_m, b_m$ are soft SUSY-breaking couplings. 
The size of the soft SUSY-breaking terms is represented by the gravitino mass $m_{3/2}$ 
as usual in the gravity mediation. 

Finally, the whole scalar potential is given by collecting the above contributions, 
\begin{align}
\label{AD/PQ_V}
V=V_{\rm Hubble}+V_{\rm soft}+V_{\rm F} + V_{\rm A}. 
\end{align}

\subsubsection{Dynamics of the AD/PQ fields during inflation}
\label{subsec:3_1_2}

In this section, we show the dynamics of the AD/PQ fields during inflation.  
We find a minimum of the scalar potential 
and consider the dynamics of the AD/PQ fields around the minimum. 
The details of the calculation are shown in Appendix~A. 

During inflation, $H \gg m_{3/2}$, the soft SUSY-breaking terms can be neglected. 
We first focus on the phase-dependent part of the potential. 
The fields can be decomposed to 
\begin{align}
 \phi_i = \hat{\phi}_i e^{i\theta_{\phi_i}},\quad \phi_i = S_0,\ S_1,\ S_2,\ \phi,   
\end{align}
where $\hat{\phi}_i$ and $\theta_{\phi_i}$ are real fields. The phase-dependent scalar potential is given by 
\begin{align}
V_{\rm phase}=& -2 \hka^2 f^2 \hSo \hSt \cos{(\theta_{S_1}+\theta_{S_2})} 
                            -2\hka \hat{\gamma} \frac{\hSz \hSo^2 \hSt  \hph^3}{M_P^3} 
          \cos{(\theta_{S_0}+\theta_{S_2}-2\theta_{S_1}-3\theta_\phi+\zeta)}  
                            \nonumber \\
                         & +2 \hka \hbh  (H \hSz) \hSo \hSt 
                           \cos{(\theta_{S_0}+\theta_{S_1}+\theta_{S_2}+\eta)}   
                           -2 \hka \hbh f^2 (H \hSz) \cos{(\theta_{S_0}+\eta)} 
                            \nonumber \\ 
                        & -\hat{\gamma} \hah H\frac{\hSo^3 \hph^3}{3M_P^3} 
                               \cos{(3 \theta_{S_1} + 3\theta_\phi+\xi)} , 
\end{align}
where the coupling constants with a hat stand for their absolute values 
and $\zeta \equiv \text{Arg}(\kappa^*\gamma)$,  
$\eta \equiv \text{Arg}(\kappa b_H)$ and $\xi \equiv \text{Arg}(\gamma a_H)$ are defined.  
The first line comes from the $F$-term potential $V_{\rm F}$ 
and the last two lines come from the Hubble-induced $A$-terms in $V_{\rm A}$. 
If $H\langle\hSz\rangle \ll f^2$, as shown later, 
the second line can be neglected and the minimum of the phases are placed at 
\begin{align}
\langle\theta_{S_1}+\theta_{S_2}\rangle&\simeq0, \nonumber \\
\langle\theta_{S_0}+\theta_{S_2}-2\theta_{S_1}-3\theta_{\phi}\rangle&\simeq -\zeta, \nonumber \\
\langle3\theta_{S_1}+3\theta_{\phi}\rangle&\simeq -\xi.
\end{align}
Under this condition, 
the extremal condition for the radial directions $\partial V/\partial \hat{\phi}_i = 0$ 
has a solution:  
\begin{align}
\langle \hSz \rangle&=
 \frac{\hka \langle\hSt\rangle}
        {\hka^2\langle\hSo\rangle^2+\hka^2 \langle\hSt\rangle^2+c_0H^2}
\frac{\hat{\gamma} \langle\hSo\rangle^2 \langle\hph \rangle^3}{M_P^3}, \nonumber \\ 
\langle \hSt\rangle&\simeq\frac{f^2}{\langle \hSo \rangle}, \quad 
\langle\hSo\rangle\simeq k\langle \hph \rangle, \nonumber \\
\langle\hph\rangle&\simeq\biggl(\frac{k\hah+\sqrt{k^6\hah^2+4c_1(2k^2+3k^4)}}{2(2k^2+3k^4)}
                                \frac{HM_P^3}{\hat{\gamma}} \biggl)^{\frac{1}{4}},
\label{eq-init}
\end{align}
where $k$ is an $\mathcal{O}(1)$ constant which depends on $\hat{a}_H, c_1, c_3$.
At this minimum,  our assumption prior to the estimation,  
\begin{align}
 H \langle\hSz\rangle \lesssim 
                 f^2 H/\hph \ll f^2,  
\end{align}
is satisfied, and the estimation is self-consistent as long as $ H \ll M_P $. 

Since the masses of AD/PQ fields $m_{\phi_i}\ (\phi_i=S_0,\ S_1,\ S_2,\ \phi)$ at the extrema are estimated by using Eqs.~(\ref{AD/PQ_V}) and (\ref{eq-init}),
\begin{align}
m_{S_0} \simeq m_{S_2} \simeq \langle \hSo \rangle \gg |m_{S_1}|, |m_{\phi}|\simeq H,
\label{AD/PQmass}
\end{align}
$S_0$ and $S_2$ are expected to have large positive mass terms and to be fixed at the minimum
during inflation. 
The curvature along the $\hSo, \hph$ directions is determined by the mass matrix, 
\begin{align}
\frac{1}{2}
    \begin{pmatrix}
   \dfrac{\partial^2 V}{\partial \hSo \partial \hSo} & 
   \dfrac{\partial^2 V}{\partial \hSo \partial \hph } \\[2.5ex]
   \dfrac{\partial^2 V}{\partial \hph \partial \hSo} & 
   \dfrac{\partial^2 V}{\partial \hph \partial \hph}
    \end{pmatrix}.
%
%
%
\label{eq:massesinf} 
\end{align}
Approximately, if $|a_H| \gg c_1,c_3$, 
the curvature is positive which is confirmed by numerical calculation. 
As long as this condition is satisfied, 
the AD field $\phi$ and the PQ field $S_1$ have masses of $\mathcal{O}(\hat{a}_H H)$.  
Therefore, all of the AD/PQ fields will settle at the minimum
during inflation. 

At the end of this subsection, we comment on the axionic isocurvature perturbation~\cite{Aiso1,Aiso2,Aiso3,Aiso4} 
and the baryonic isocurvature perturbation~\cite{Biso1,Biso2,Biso3,Biso4,Biso5}.
If there is no Hubble-induced $A$-term with $a_HH$, the phase direction $3\theta_{S_1}+3\theta_{\phi}+\xi$ becomes massless. Then the large baryonic isocurvature perturbation is induced. In our model, the phase direction $3\theta_{S_1}+3\theta_{\phi}+\xi$ has a mass of $\mathcal{O}(H)$. This avoids the problematic baryonic isocurvature perturbation.
The axionic isocurvature perturbation will also be so suppressed by the large VEVs of PQ and AD fields that the model is consistent 
with the Planck observations~\cite{Planck2018}.\footnote{For more recent works on axion 
isocurvature perturbations, see e.g., Refs.~\cite{Kadota:2015uia,Schmitz:2018nhb}.} 
The axionic isocurvature perturbation is discussed in more detail in Sec.~\ref{subsec:iso}.

\subsubsection{Dynamics of the AD/PQ fields after inflation: $H>m_{\phi}$}
\label{subsec:3_1_3}

After the end of inflation, the inflaton starts to oscillate around $I_{\rm min}$ 
and the effect of Hubble-induced $A$-term $a_H H$ on the dynamics of the AD/PQ fields turns off. 
In this era, the Universe is dominated by the oscillation energy 
and the Hubble parameter evolves as $H = 2/(3t)$. 
The K$\ddot{\rm a}$hler potential and superpotential are approximately 
\begin{align}
K_{\rm inf}=& |I|^2+\cdots 
                =I_{\rm min}^{\ast} \dI +I_{\rm min} \dI^{\dag}+|\dI|^2+\mathcal{O}(\dI^3), 
                \nonumber \\
W=& \frac{1}{2}M_{\rm inf}(I-I_{\rm min})^2+...
         =\frac{1}{2}M_{\rm inf}\dI^2+\mathcal{O}(\dI^3),
\end{align}
where $\dI = I-I_{\rm min}$, and $M_{\rm inf}$ is the inflaton mass. 
Since $F_I = -M_{\rm inf} \dI^\dagger$ 
and the Hubble-induced $A$-term $a_H H$ originates from $F_I$, 
$a_H$ diminishes rapidly after the $F$-term inflation~\cite{Biso5,ADmulti} as long as the inflaton oscillates in the period $M_{\rm inf}^{-1}$, which is shorter 
than the Hubble time $H^{-1}$.  
The suppression factor is estimated as~\cite{ADmulti}
\begin{align}
\frac{H}{H_{\rm inf}},
\end{align}
where $H_{\rm inf}$ is the Hubble parameter during inflation. 
The Hubble-induced $A$-term with $a_HH$ after inflation is given by
\begin{align}
\frac{\gamma a_HH^2S_1^3\phi^3}{3H_{\rm inf}M_P^3}+ {\rm H.c.}.
\end{align}
Eventually, this Hubble-induced $A$-term diminishes as time evolves.   
The point in Eq.~(\ref{eq-init}) is no longer the minimum of the potential   
and the value of this extrema changes. Then the AD/PQ fields start to roll down.
Since Eqs.~(\ref{AD/PQ_V}),(\ref{eq-init}), and~(\ref{AD/PQmass}) show that $S_0$ and $S_2$ are heavier than $S_1$ and $\phi$ until $\langle \hSo \rangle$ is larger than $f$,
we assume that $S_0$ and $S_2$ are fixed at the minimum in this epoch. 
The dynamics of $S_1$, $\phi$ obey the following equations of motion: 
\begin{align}
\frac{d^2S_1}{dt^2}+\frac{2}{t}\frac{dS_1}{dt}+\frac{\partial V}{\partial S_1^{\dag}}=0, \nonumber \\
\frac{d^2\phi}{dt^2}+\frac{2}{t}\frac{d\phi}{dt}+\frac{\partial V}{\partial \phi^{\dag}}=0.
\label{EOM_H>m}
\end{align}
The soft SUSY-breaking terms can also be neglected 
until $H \sim m_\phi \sim m_{3/2}$. 

\begin{figure}[htbp]
  \begin{minipage}{0.5\hsize}
   \begin{center}
     \includegraphics[clip,width=80mm]{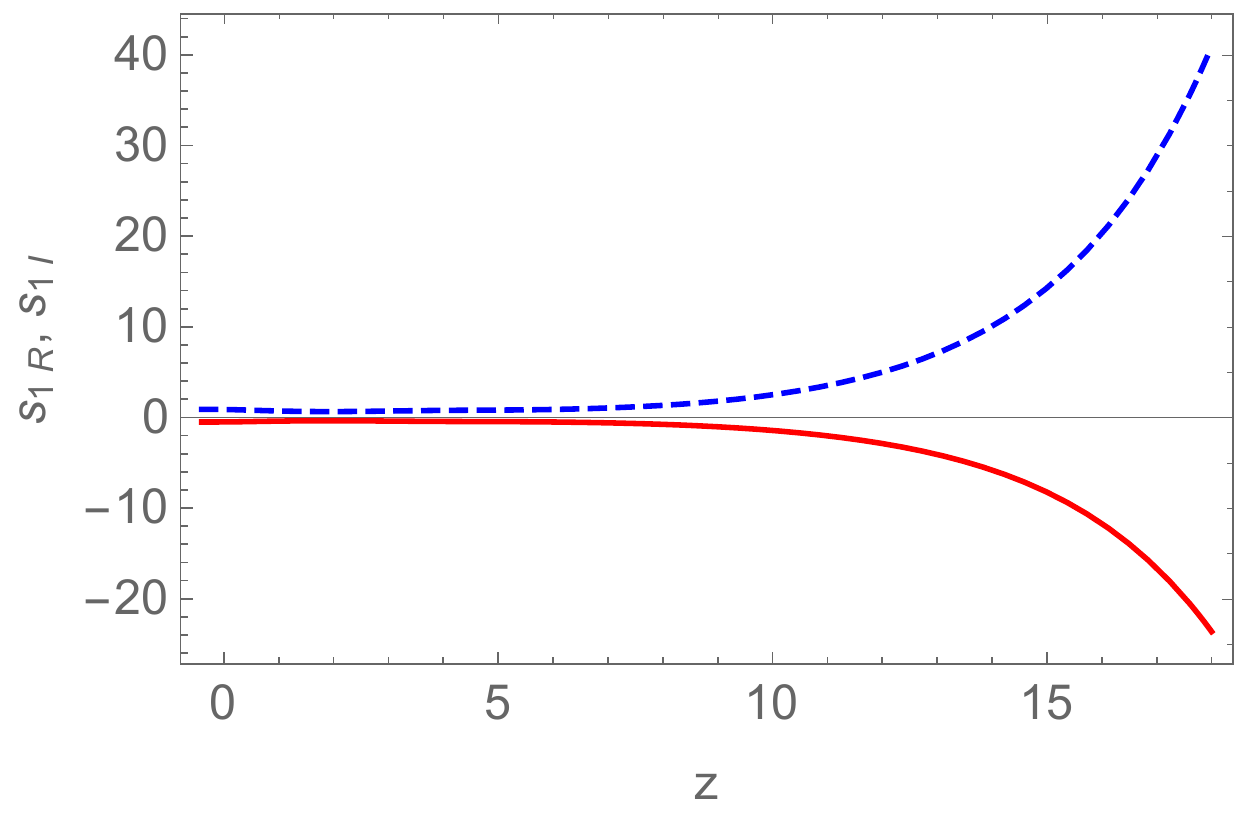}
    \end{center}
   \end{minipage}
  \begin{minipage}{0.5\hsize}
   \begin{center}
    \includegraphics[width=80mm]{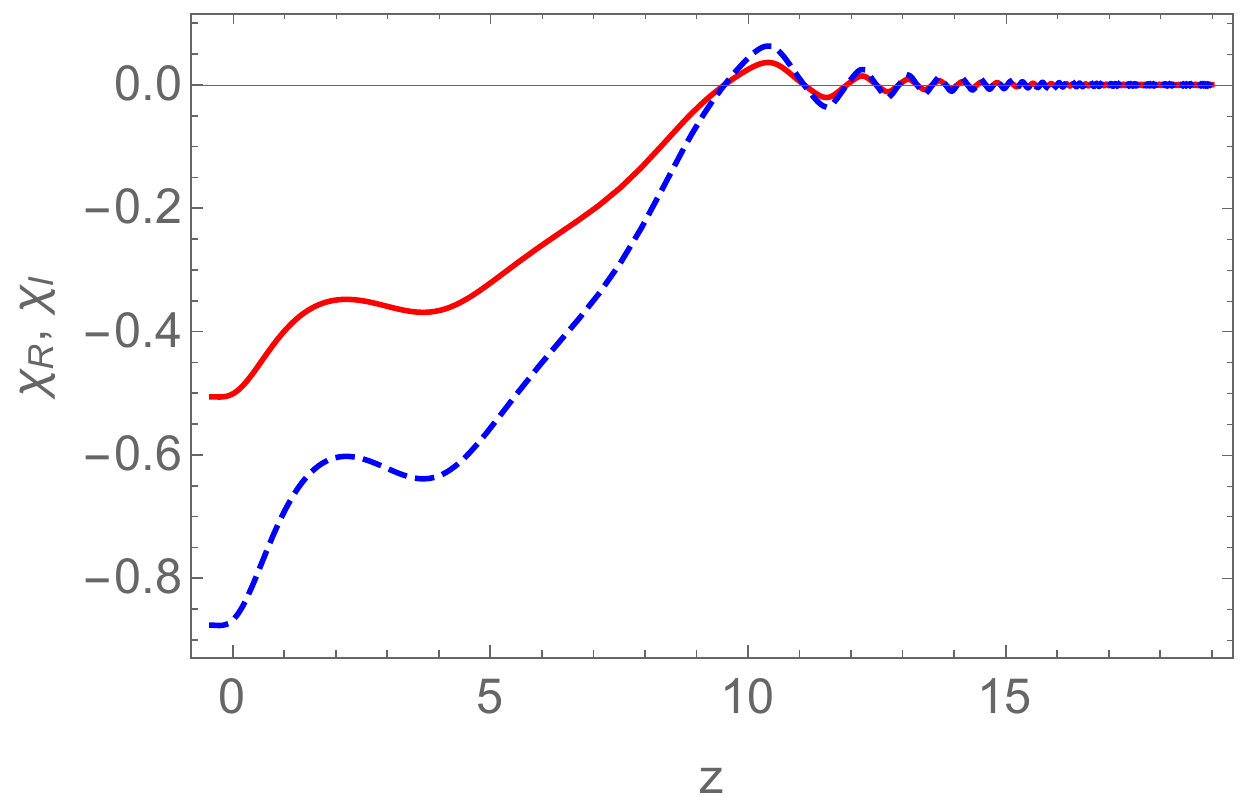}
   \end{center}
  \end{minipage}
 \caption{The time variations of $s_1, \chi$  for $c_1> c_3$ during $m_{\phi}<H$. In the left panel, we draw the time variations of $(s_{1R},s_{1I})$, whereas, in the right panel, we draw the time variations of $(\chi_R, \chi_I)$. The red (blue dashed) line represents the real (imaginary) part of the scalar field. We set the parameters as $\hat{\gamma}=1, \hat{a}_H=5,c_1=1/4,c_3=1/5$, and $\xi=0$, and the initial conditions as $\theta_{s_1}|_{t=t_{\rm end}}=\frac{2\pi}{3}, \theta_{\phi}|_{t=t_{\rm end}}=-\frac{2\pi}{3}$ and $\frac{ds_1}{dt}|_{t=t_{\rm end}}=0, \frac{d\chi}{dt}|_{t=t_{\rm end}}=0$.\ We draw these time variations from $H_{\rm inf}\simeq 10^{11}~{\rm GeV}(z\simeq\log(2/3))$ to $H\simeq m_{\phi}\simeq 1~{\rm TeV}(z\simeq 18)$.} 
 \label{fig:noAterm}
 \end{figure}

We numerically solve the equations. 
Let us introduce new parameters $z$, $s_1$, and $\chi$, defined as 
\begin{align}
z=\log H_{\rm inf}t,\quad 
S_1= s_1 \biggl(\frac{2H_{\rm inf}M_P^3}{3\hat{\gamma}}e^{-z} \biggl)^{\frac{1}{4}}, 
 \quad 
\phi=\chi \biggl(\frac{2H_{\rm inf}M_P^3}{3\hat{\gamma}}e^{-z} \biggl)^{\frac{1}{4}}, 
\end{align}
from which Eq.~(\ref{EOM_H>m}) becomes
\begin{align}
\frac{\partial^2 s_1}{\partial z^2}+\frac{1}{2}\frac{\partial s_1}{\partial z}
-\biggl(\frac{4c_1}{9}+\frac{3}{16}\biggl)s_1+\frac{8}{9}|s_1|^2|\chi|^6s_1+\frac{4}{3}|s_1|^4|\chi|^4s_1
-\frac{8\hat{a}_H}{27}e^{-z-i\xi}s_1^{\dag2}\chi^{\dag3}&=0, \nonumber \\
\frac{\partial^2 \chi}{\partial z^2}+\frac{1}{2}\frac{\partial \chi}{\partial z}
-\biggl(\frac{4c_3}{9}+\frac{3}{16}\biggl)\chi+\frac{4}{3}|s_1|^4|\chi|^4\chi
+\frac{8}{9}|s_1|^6|\chi|^2\chi-\frac{8\hat{a}_H}{27}e^{-z-i\xi}s_1^{\dag3}\chi^{\dag2}&=0.
\end{align}
The time variations of $s_1$ (with $s_{1R}={\rm Re}(s_1), s_{1I}={\rm Im}(s_1)$)  
and $\chi$ (with $\chi_R={\rm Re}(\chi), \chi_I={\rm Im}(\chi)$) 
are shown in the left panel and the right panel of Fig.~\ref{fig:noAterm}, respectively. The red (blue dashed) line represents the real (imaginary) part of scalar field. 
For concreteness,
$\hat{\gamma}=1, \hat{a}_H=5,c_1=1/4,c_3=1/5, \xi=0$ are assumed. Then, $k$ is estimated as $k \simeq 1$ for $\hat{a}_H \gg c_1, c_3$.
Since the variations of field values depend on the Hubble parameter during inflation $H_{\rm inf}$, we assume that the Hubble parameter during inflation is fixed at $H_{\rm inf} = 10^{11}$ GeV, which is consistent with the isocurvature perturbation discussed in Sec.~\ref{subsec:iso}.  
The initial conditions at inflation end $t = t_{\rm end} = H_{\rm inf}^{-1}(z=\log(2/3))$ are  
\begin{align}
\theta_{s_1}|_{z=\log(2/3)}=\frac{2\pi}{3}, \quad 
\theta_{\phi}|_{z=\log(2/3)}=-\frac{2\pi}{3}
\end{align}
and 
\begin{align}
 \frac{\partial s_1}{\partial z}\biggl |_{z=\log(2/3)}=0,\quad \frac{\partial \chi}{\partial z}\biggl|_{z=\log(2/3)}=0.
\end{align}
In addition, the initial conditions of $|s_1|, |\chi|$ are extracted from Eq.~(\ref{eq-init}),
\begin{align}
|s_1||_{z=\log(2/3)}=k|\chi||_{z=\log(2/3)}, \ \ \ \ |\chi||_{z=\log(2/3)}=\biggl(\frac{k\hah+\sqrt{k^6\hah^2+4c_1(2k^2+3k^4)}}{2(2k^2+3k^4)} \biggl)^{\frac{1}{4}}.
\end{align}
Hereafter we take $k=1$.

The trajectories start from $t = H_{\rm inf}^{-1}\ (z=\log(2/3))$ 
and evolve to $t = m_{\phi}^{-1} \simeq 1~{\rm TeV^{-1}}\ (z\simeq18)$. 
Finally, $s_1,\chi$ at $H\simeq m_{\phi} \simeq 1~{\rm TeV}(z\simeq18)$ 
have the following values, 
\begin{align}
s_{1R}&\simeq -23.6,\ \  s_{1I}\simeq40.9,\ \ \chi_R\simeq -7.17 \times 10^{-3},\ \chi_I\simeq -1.24 \times 10^{-3}, \nonumber \\
 \partial_z{s}_{1R}&\simeq -8.29,\ \partial_z{s}_{1I}\simeq 14.4,\ \partial_z{\chi}_R\simeq 1.14\times 10^{-2},\ \partial_z{\chi}_I\simeq 1.97\times 10^{-2}.
 \label{ini_osc}
\end{align} 
These values give initial values for the dynamics at $H \simeq m_\phi$ 
which is analyzed in the next subsection. 
The numerical calculation indicates 
that $|s_1|$ increases by $\mathcal{O}(10)$, while $|\chi|$ diminishes by $\mathcal{O}(10^3)$ as the time evolution for $c_1>c_3$.
This result will not depend on the parameters or the initial values, except for $c_1,c_3$ sensitively. However, the variations due to the dynamics will decrease for smaller $H_{\rm inf}$ and $c_1, c_3$. 
Note that $S_1$ does not dominate the energy density of the Universe. Although $|s_1|$ increases during the time $m_{\phi} \lesssim H$, the field value of $\hat{S_1}$ is still smaller than the reduced Planck mass,
\begin{align}
\hat{S_1} \ll M_P.
\end{align}
Thus, the inflaton dominates the Universe during $m_{\phi} \lesssim H$,
\begin{align}
\rho_{S_1} \sim H^2\hat{S}_1^2 \ll \rho_{\rm inf} \sim 3H^2M_P^2.
\end{align}

\begin{figure}[htbp]
  \begin{minipage}{0.5\hsize}
   \begin{center}
     \includegraphics[clip,width=80mm]{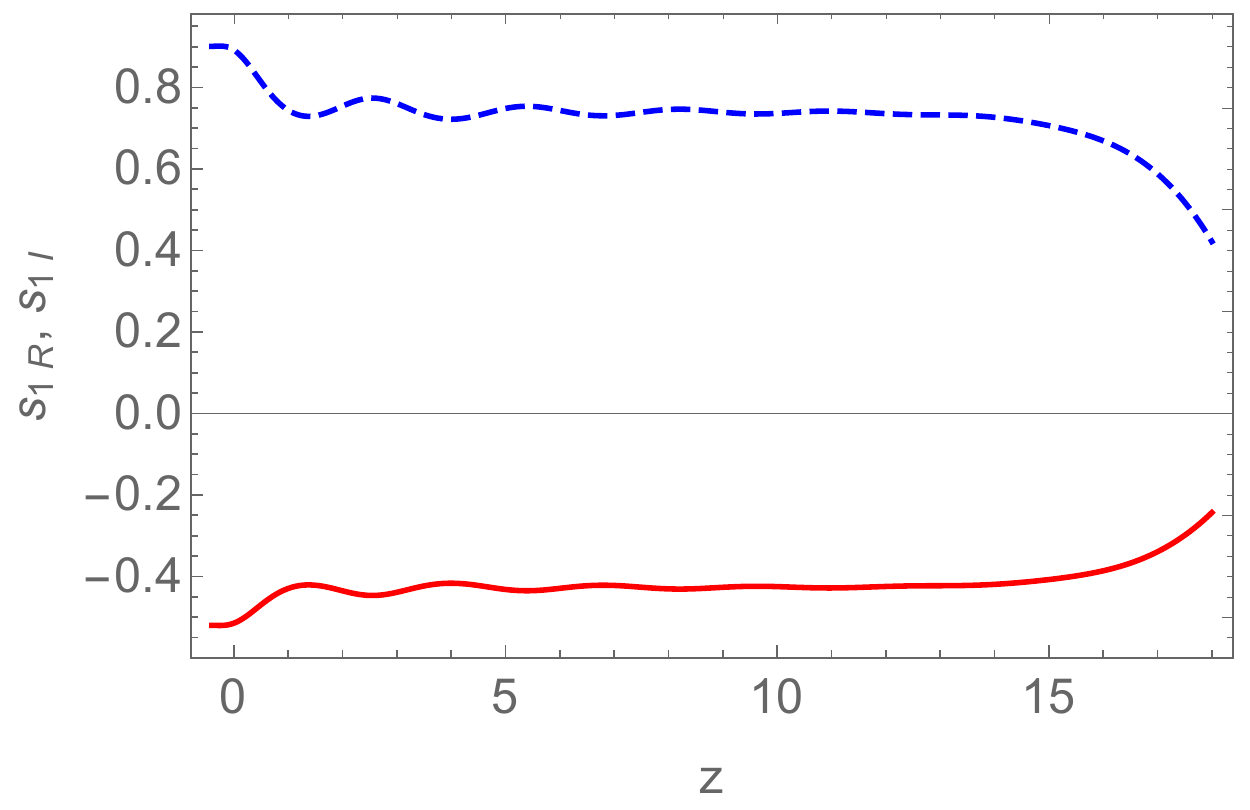}
    \end{center}
   \end{minipage}
  \begin{minipage}{0.5\hsize}
   \begin{center}
    \includegraphics[width=80mm]{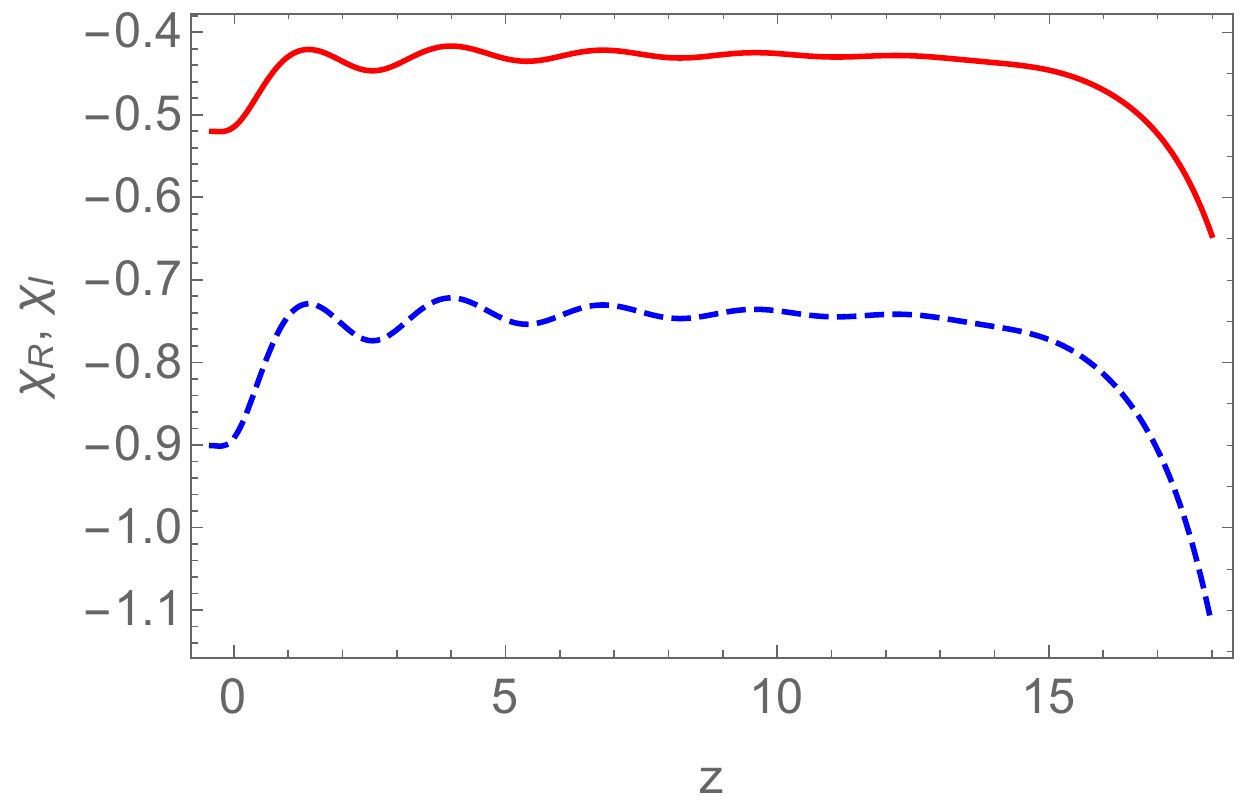}
   \end{center}
  \end{minipage}
 \caption{The time variations of $s_1, \chi$  for $c_1= c_3$ during $m_{\phi}<H$. In the left panel, we draw the time variations of $(s_{1R},s_{1I})$, whereas, in the right panel, we draw the time variations of $(\chi_R, \chi_I)$. The red (blue dashed) line represents the real (imaginary) part of the scalar field. We set $c_1=c_3=1$ and $\xi=0$. The values of the other parameters and the initial conditions are the same as in Fig.~\ref{fig:noAterm}.\ We draw these time variations from $H_{\rm inf}\simeq 10^{11}~{\rm GeV}(z\simeq\log(2/3))$ to $H\simeq m_{\phi}\simeq 1~{\rm TeV}(z\simeq 18)$.} 
 \label{fig:noAterm2}
 \end{figure}

In addition, the dynamics of $|s_1|$ and $|\chi|$ are characteristic for the $c_1=c_3$ case. Although the point in Eq.~(\ref{eq-init}) becomes a saddle point after inflation, $|s_1|$ and $|\chi|$ oscillate around the ridges of the saddle point. Because of the large field values of $\hat{S_1}$ and $\phi$, the quantum fluctuations of $\hat{S_1}$ and $\phi$ will not affect the dynamics of $|s_1|$ and $|\chi|$ at the time $m_{\phi}\lesssim H$. Fig.~\ref{fig:noAterm2} shows the time variations of $s_1$ and $\chi$ for $c_1=c_3=1$. The values of the other parameters and the initial conditions are the same as in Fig.~\ref{fig:noAterm}. From this perspective, we confirm that both $|s_1|$ and $|\chi|$ stay around the $\mathcal{O}(1)$ values.

In the $c_1<c_3$ case, we find that $|s_1|$ decreases while $|\chi|$ increases in typical parameter spaces. 
Since we must consider the dynamics of $S_2$ for small $|s_1|$, the dynamics of AD/PQ fields after inflation are complicated 
and we leave them to a future work. 
In the next section, we therefore concentrate on two cases $c_1 > c_3$ and $c_1 = c_3$, by estimating 
the dynamics of $S_1,\phi$ for $H\simeq m_{\phi}$.

\subsubsection{Dynamics of the AD/PQ fields at $H\simeq m_{\phi}$ and baryon asymmetry} 
\label{subsec:3_1_4}

In this subsection, 
we consider the dynamics of the AD/PQ fields at $H\simeq m_{3/2}$ and finally estimate the amount of baryon asymmetry.
At this epoch, 
the soft supersymmetry-breaking effect becomes important.  
The AD/PQ fields eventually start to oscillate around the minimum: 
\begin{align}
 &\langle \hat{S_0} \rangle =  \frac{\hka \langle\hSt\rangle}
        {\hka^2\langle\hSo\rangle^2+\hka^2 \langle\hSt\rangle^2+c_0H^2}
\frac{\hat{\gamma} \langle\hSo\rangle^2 \langle\hph \rangle^3}{M_P^3}=0, \nonumber \\ 
&\langle \hat{S_1} \rangle \simeq f,\ 
 \langle \hat{S_2} \rangle \simeq f,\ \langle \hat{\phi} \rangle =0.
 \label{minimum_Hm}
\end{align}
Then $S_0$ and the scalar field orthogonal to the flat direction $S_1S_2=f^2$ have masses of order of $f$. These masses are heavier than the scale of soft mass $\mathcal{O}(m_{3/2})$ which is the same order of $m_\phi$. Thus, following the previous section, we can set $S_1S_2=f^2$ and $\hat{S}_0$ as Eq.~(\ref{eq-init}).
In the following, we consider only the dynamics of $S_1$ and $\phi$ in this epoch. Since the masses of $S_1$ and $\phi$ are positive, $S_1$ and $\phi$ will go to the minimum (Eq.~(\ref{minimum_Hm})).
Because $S_1$ and $\phi$ will be damped enough, the potential of the AD field will be approximated as a quadratic one which does not depend on the PQ fields. Thus the amount of baryon asymmetry will be conserved after $S_1 \simeq f$ because we neglect CP-violating terms. 
The above statement is numerically confirmed in the following analysis. 

First, let us analytically consider the dynamics of $S_1, \phi$ and estimate the amount of baryon asymmetry.
The baryon number density is given by
\begin{align}
n_B=\frac{i}{3}\left(\frac{d\phi^{\ast}}{dt}\phi-\phi^{\ast}\frac{d\phi}{dt}\right)=\frac{2}{3}|\phi|^2\frac{d\theta_{\phi}}{dt}. 
\end{align}
It obeys the following equation of motion using the one of $\phi$,
\begin{align}
\frac{dn_B}{dt}+3Hn_B={\rm Im}\biggl(\frac{\partial V}{\partial \phi}\phi \biggl), 
\end{align}
and in the integral form, we obtain
\begin{align}
R(t)^3n_B(t)=&\int^t_{t_{\rm inf}}dt' R(t')^3{\rm Im}
    \biggl(\frac{\gamma a_mm_{3/2}S_1^3\phi^3}{3M_P^3}\biggl) \nonumber \\
    =&\int^{t_{\rm osc}}_{t_{\rm inf}}dt' R(t')^3{\rm Im}
    \biggl(\frac{\gamma a_mm_{3/2}S_1^3\phi^3}{3M_P^3}\biggl)+\int^t_{t_{\rm osc}}dt' R(t')^3{\rm Im}
    \biggl(\frac{\gamma a_mm_{3/2}S_1^3\phi^3}{3M_P^3}\biggl),
\label{eq-nint}
\end{align}
where $R$ is the scale factor of the Universe, and 
$t_{\rm inf} \sim H_{\rm inf}^{-1}$ is the time at the end of inflation. 
$\phi$ starts to oscillate after the time $t_{\rm osc}$ defined as $H_{\rm osc}=2/(3t_{\rm osc})\sim m_{\rm \phi}$.
The CP-violating factor is defined as $\delta_{\rm eff}=\sin(\xi+3\theta_{S_1}+2\theta_{\phi})$. 
As explicitly checked in the numerical calculation, the second integration on the second line of Eq.~(51) gives small effects to
the baryon asymmetry since the sign phase factor $\delta_{\rm eff}$ changes rapidly after the AD/PQ fields $\phi,S_1$ start to oscillate. 
As a result, the baryon number in Eq.~(\ref{eq-nint}) will be fixed at $t=t_{\rm osc}$ and the baryon number density at $t=t_{\rm osc}$ is estimated as
\begin{align}
n_B(t_{\rm osc})\simeq \frac{1}{3}\epsilon \hat{a}_mm_{3/2}\delta_{\rm eff}\biggl(\frac{m_{3/2}M_P^3}{\hat{\gamma}} \biggl)^{\frac{1}{2}},
\label{LD}
\end{align} 
where $\epsilon$ is defined as
\begin{align}
\epsilon=\hat{S_1}(t_{\rm osc})^3\hat{\phi}(t_{\rm osc})^3\biggl(\frac{m_{3/2}M_P^3}{\hat{\gamma}} \biggl)^{-\frac{3}{2}}.
\end{align}

To check the above statements, we numerically solve the equations of motion of $S_1,\phi$ and estimate the trajectories and the baryon asymmetry for $H \lesssim m_{\phi}$. Here we assume that the inflaton still dominates the Universe, namely the matter-dominated Universe. Reparametrizing again as
\begin{align}
S_1&\rightarrow S_1 \biggl(\frac{m_{3/2}M_P^3}{\hat{\gamma}}\biggl)^{\frac{1}{4}},\nonumber \\
\phi&\rightarrow \phi \biggl(\frac{m_{3/2}M_P^3}{\hat{\gamma}} \biggl)^{\frac{1}{4}},
\end{align}
Eq.~(\ref{EOM_H>m}) approximately becomes
\begin{align}
\frac{d^2S_1}{dt^2}+\frac{2}{t}\frac{dS_1}{dt}+2m_{3/2}^2\hat{S_1}^2\hat{\phi}^6S_1+3m_{3/2}^2\hat{S}_1^4\hat{\phi}^4S_1-a_mm_{3/2}^2S_1^{\dag2}\phi^{\dag3} \nonumber \\ 
+\biggl(m_1^2-\frac{4c_1}{9t^2}\biggl)S_1-\biggl(m_2^2+\frac{4c_2}{9t^2} \biggl)\frac{f^4}{m_{3/2}M_P^{3}\hat{S}_1^4}S_1&=0, \nonumber \\
\frac{d^2\phi}{dt^2}+\frac{2}{t}\frac{d\phi}{dt}+3m_{3/2}^2\hat{S}_1^4\hat{\phi}^4\phi+2m_{3/2}^2\hat{S_1}^6\hat{\phi}^2\phi-a_mm_{3/2}^2S_1^{\dag3}\phi^{\dag2} 
+\biggl(m_{\phi}^2-\frac{4c_3}{9t^2} \biggl)\phi&=0,
\label{EOM_Hm}
\end{align}
from which the oscillation time is extracted as $t_{\rm osc}=2c_3^{1/2}/(3 m_\phi)$. Here we set $m_1=m_2=m_{3/2}, \kappa=\gamma=1$ for simplicity.
In the following analysis, we numerically solve Eq.~(\ref{EOM_Hm}) from the time $t_{\rm osc}$. 

\begin{figure}[htbp]
  \begin{minipage}{0.5\hsize}
   \begin{center}
     \includegraphics[clip,width=70mm]{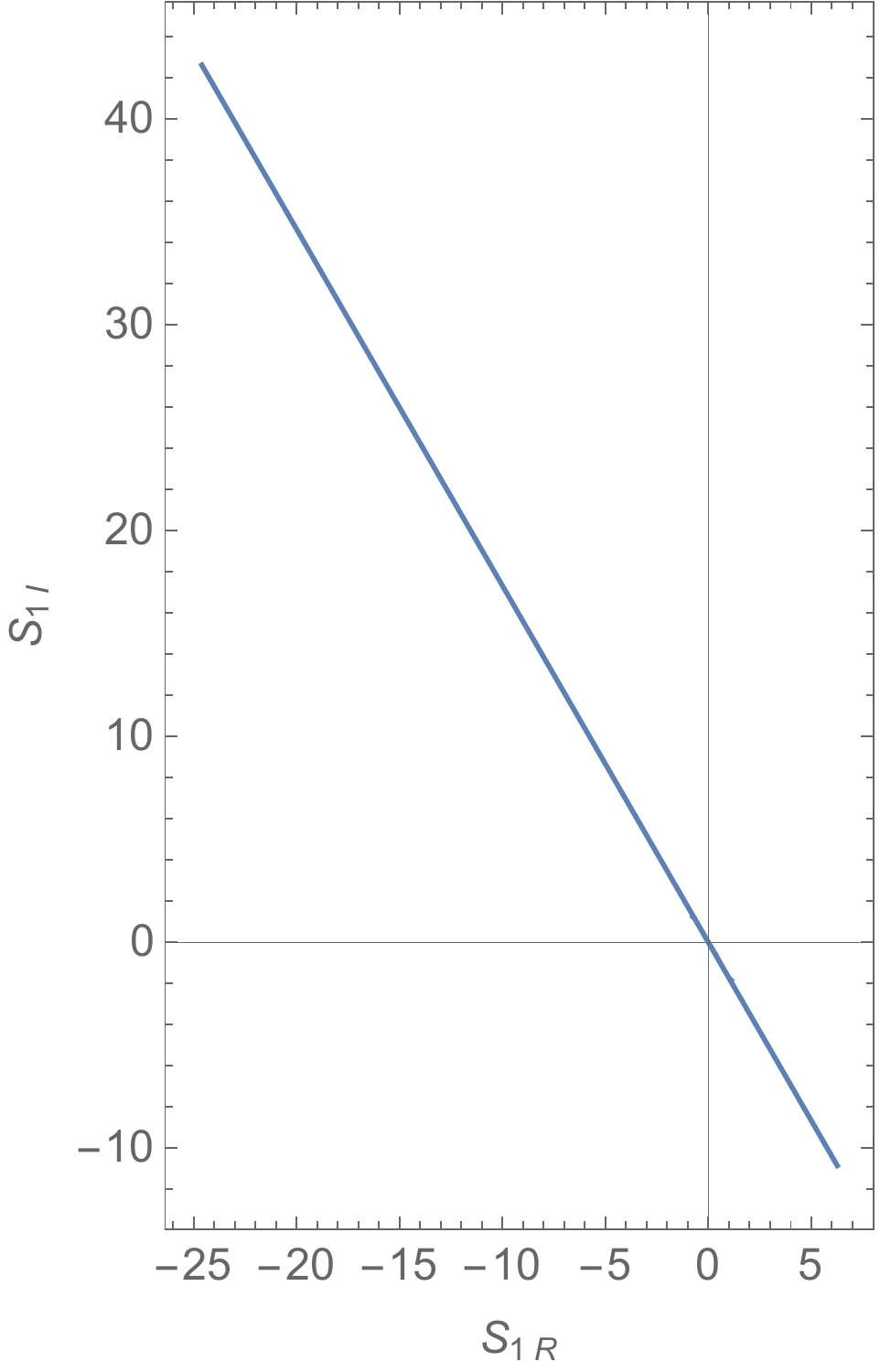}
    \end{center}
   \end{minipage}
  \begin{minipage}{0.5\hsize}
   \begin{center}
    \includegraphics[width=70mm]{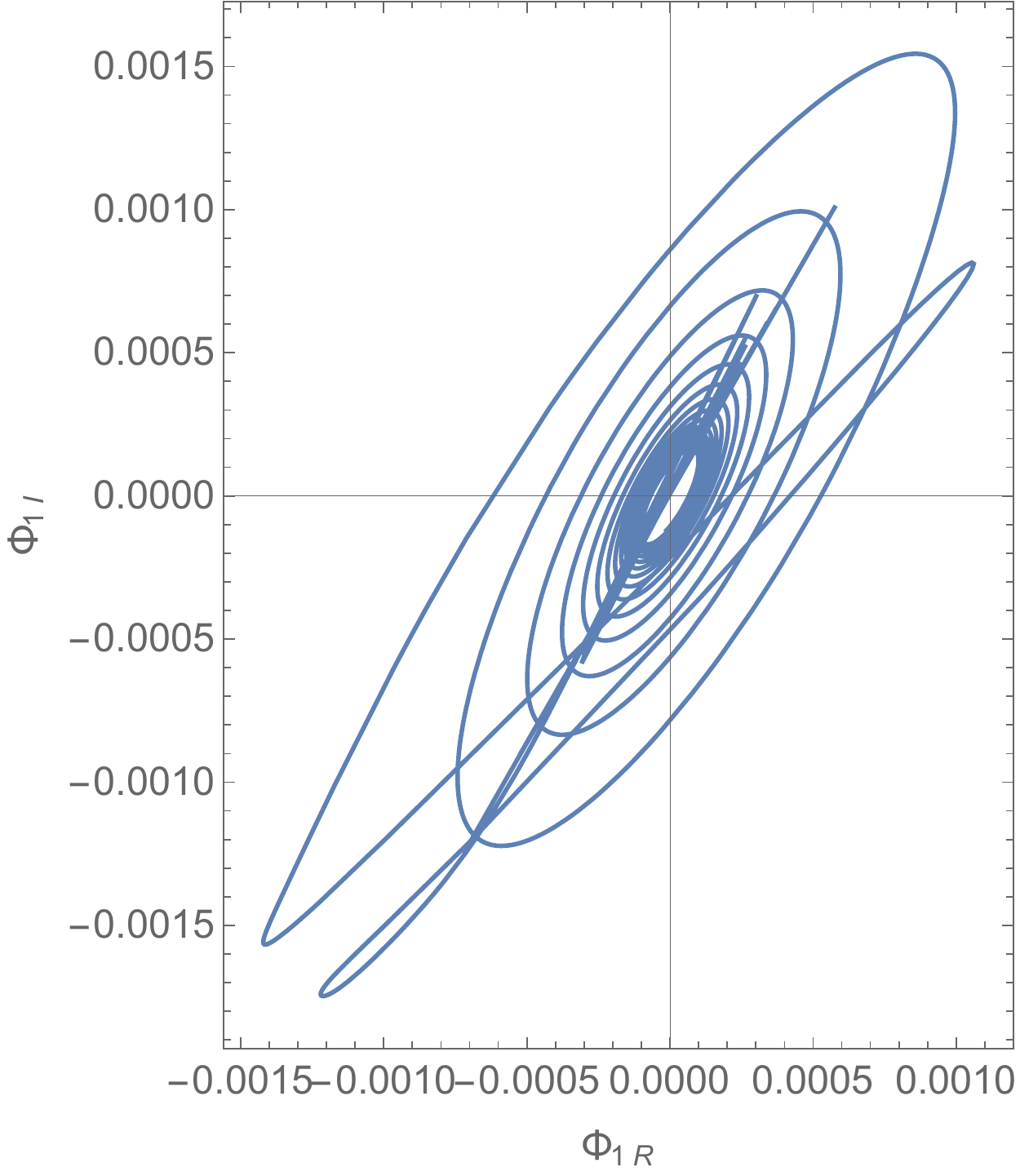}
   \end{center}
  \end{minipage}
 \caption{The dynamics of $S_1, \phi$ for $c_1>c_3$ during $H<m_{\phi}$. 
 In the left (right) panel, we draw the trajectories of $(S_{1R}, S_{1I})$ ($(\phi_{R}, \phi_{I})$) as a function of $z$. 
We set the parameters as $\hat{\gamma}=1, \hat{a}_m=c_2=1,c_1=1/4, c_3=1/5, f=10^{9}$ in units of $m_{3/2}=10^3~{\rm GeV}=1$, 
$\arg(\gamma a_m)=\pi/3$ and the initial conditions as in Eq.~(\ref{ini_osc}).}
 \label{fig:oscillate}
 \end{figure}
\begin{figure}[htbp]
   \begin{center}
     \includegraphics[clip,width=8.0cm]{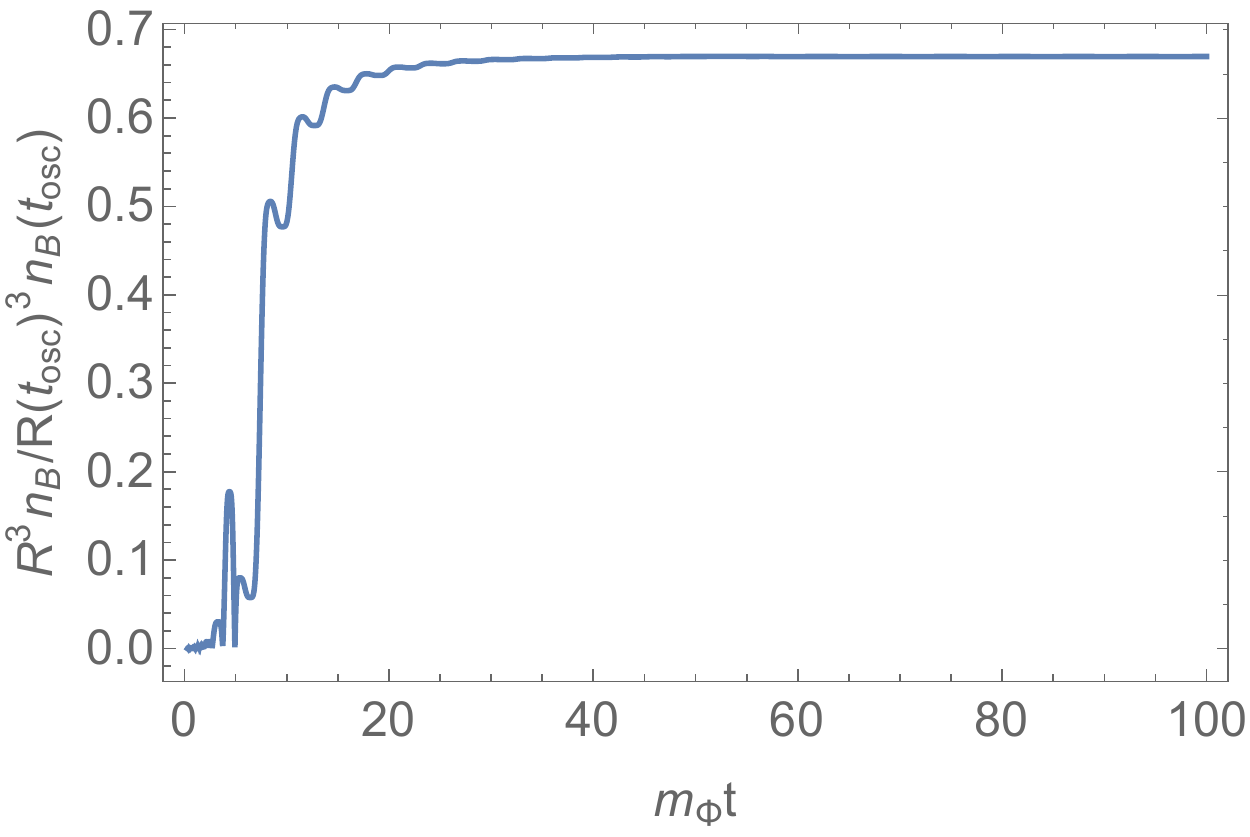}
    \end{center}
 \caption{The time evolution of the ratio between the numerical value $R(t)^3n(t)$~(\ref{eq-nint}) and $R^3(t_{\rm osc})n(t_{\rm osc})$~(\ref{LD}) for $c_1 > c_3$. The horizontal axis corresponds to $m_{\phi}t$. Here we set $\epsilon=3.1 \times 10^{-4}$ and $\delta_{\rm eff}=1$. The parameters are the same as in Fig.~\ref{fig:oscillate}. }
 \label{fig:BA}
 \end{figure}

First, let us consider the case with $c_1>c_3$.  We solve Eq.~(\ref{EOM_Hm}) and draw the trajectories of  $(S_{1R}={\rm Re}(S_1), S_{1I}={\rm Im}(S_1))$ in the right panel of Fig.\ref{fig:oscillate} and the trajectories of $(\phi_R={\rm Re}(\phi), \phi_I={\rm Re}(\phi))$ in the left panel of Fig.~\ref{fig:oscillate}. Here we set the parameters as $m_{\phi}=1, \hat{a}_m=c_2=1,c_1=1/4, c_3=1/5, \arg(a_m)=\pi/3$, $f=10^9$ in units of $m_{3/2}=10^3~{\rm GeV}=1$ and the initial conditions as in Eq.~(\ref{ini_osc}). We take the initial time as $t_{\rm osc}=2c_3^{1/2}/(3m_{\phi})$. Fig.~\ref{fig:oscillate} shows that the AD field $\phi$ and the PQ field $S_1$ go to the minimum (\ref{minimum_Hm}). 
Then we can estimate $\epsilon \simeq 3.1 \times 10^{-4}$ numerically, and the ratio of the numerical value in Eq.~(\ref{eq-nint}) to the analytical value in Eq.~(\ref{LD}) is shown in Fig.~\ref{fig:BA}.  It turns out that the numerical value coincides with the analytical one in Eq.~(\ref{LD}) with $\delta_{\rm eff}=1$ after the AD/PQ fields start to oscillate. 
However, we find that the numerical value is a little smaller than the analytical estimation and the ratio of the numerical value to the analytical one becomes $\mathcal{O}(10)$ in other parameter regions. This is because the rotation of $\phi$ is complicated as in Fig.~\ref{fig:oscillate}, and then the baryon number in Eq.~(\ref{eq-nint}) is not exactly fixed at $t_{\rm osc}$. 
The mass of $\theta_\phi$ is heavier than the mass of $\theta_{S_1}$ because in Fig.~\ref{fig:noAterm}, the field value of $\hat{\phi}$ decreases while $\hat{S}_1$ increases and the field value of $\hat{\phi}$ is smaller than the one of $S_1$. 
For this reason, $\theta_\phi$ moves whereas $\theta_{S_1}$ does not as shown in Fig.~\ref{fig:oscillate}.

\begin{figure}[htbp]
  \begin{minipage}{0.5\hsize}
   \begin{center}
     \includegraphics[clip,width=7.0cm]{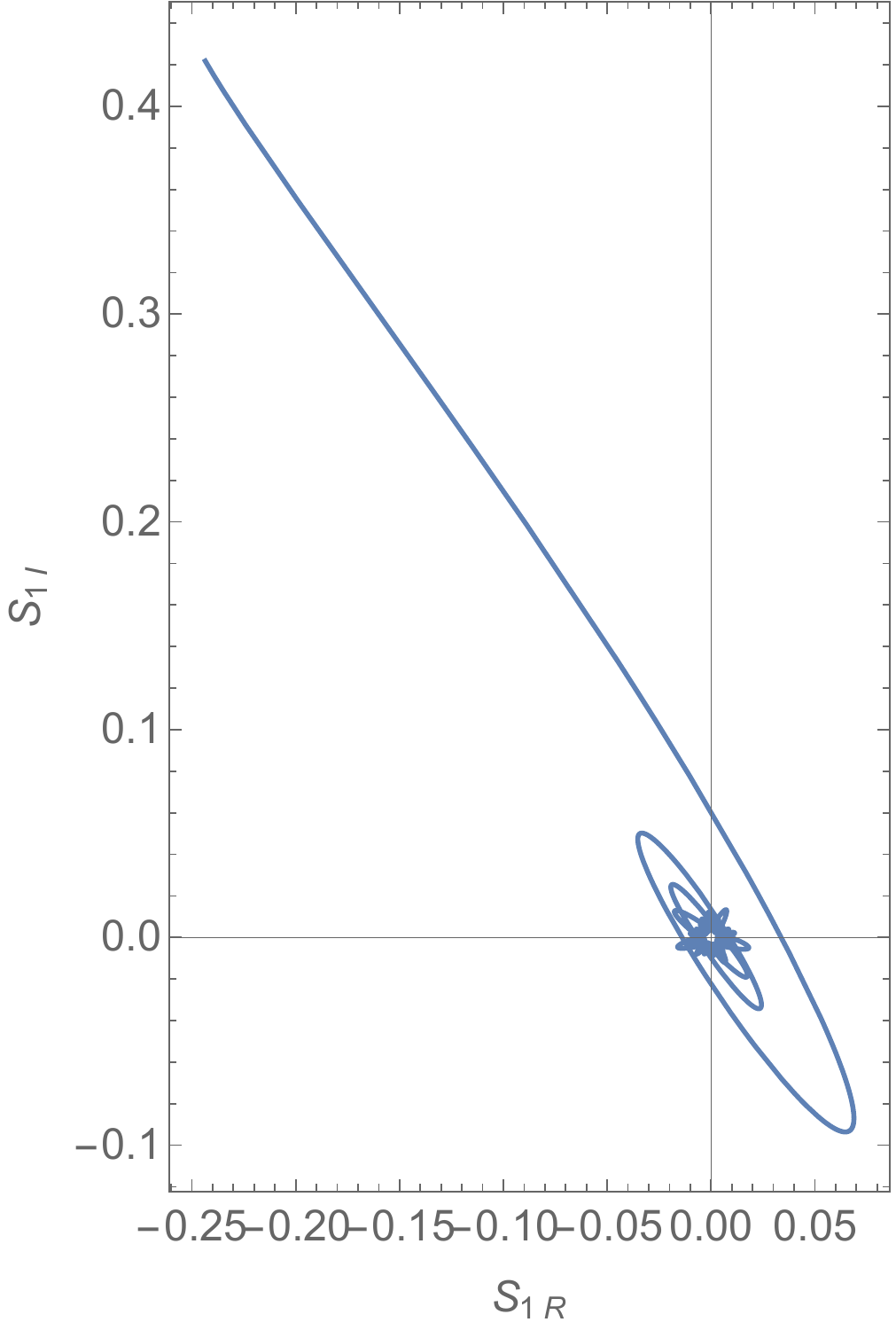}
    \end{center}
   \end{minipage}
  \begin{minipage}{0.5\hsize}
   \begin{center}
    \includegraphics[width=70mm]{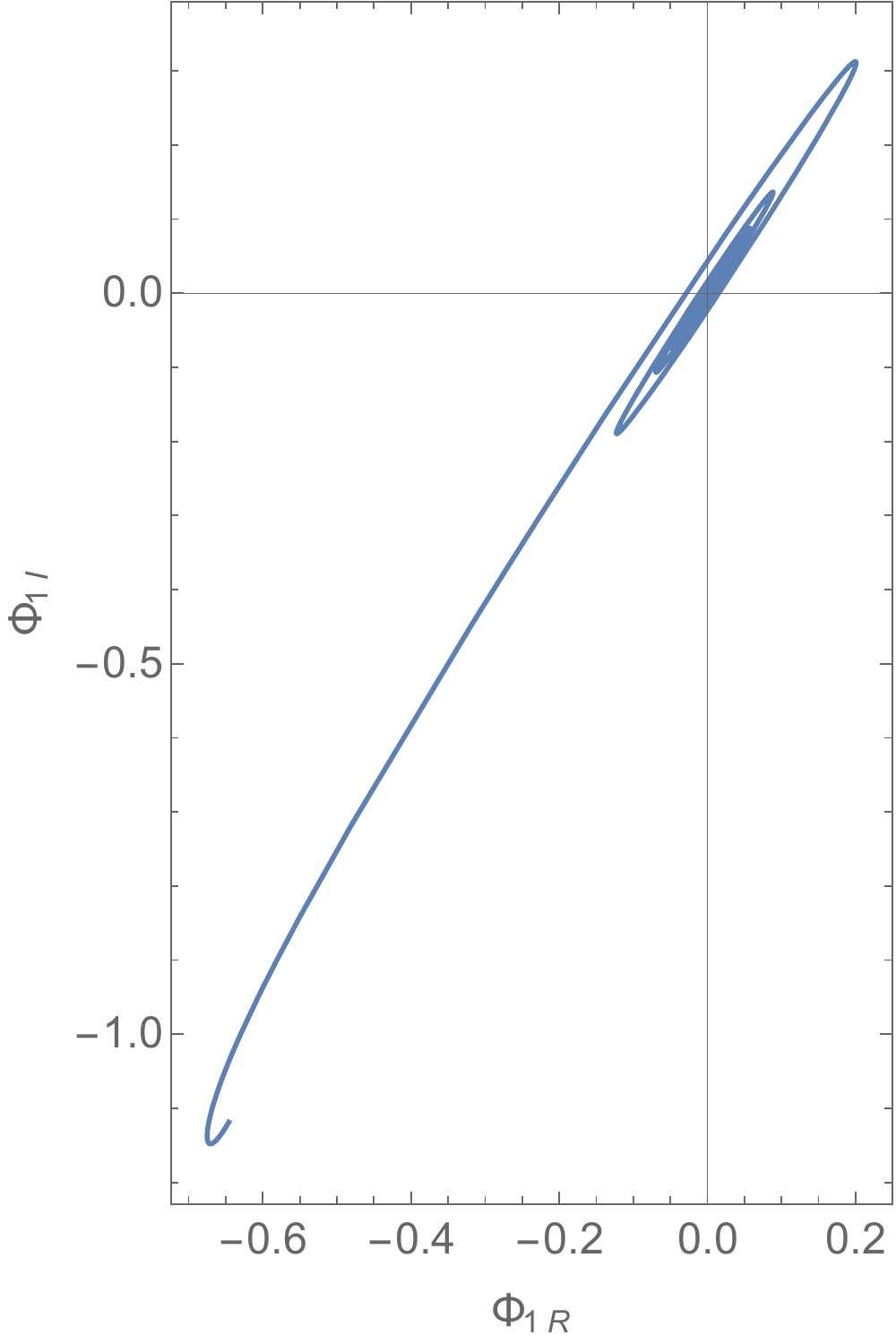}
   \end{center}
  \end{minipage}
 \caption{The dynamics of $S_1, \phi$ for $c_1=c_3$ during $H<m_{\phi}$. 
  In the left (right) panel, we draw the trajectories of $(S_{1R}, S_{1I})$ ($(\phi_{R}, \phi_{I})$) as a function of $z$. 
  We set the parameters as $\hat{\gamma}=\hat{a}_m=c_1=c_2=c_3=1, f=10^{9}$ in units of $m_{3/2}=10^3~{\rm GeV}=1$, 
  $\arg(\gamma a_m)=\pi/3$ and the initial conditions are the values of $s_1,\chi$ at $z=18$ in Fig.~\ref{fig:noAterm2}.}
 \label{fig:oscillate_2}
 \end{figure}

\begin{figure}[htbp]
   \begin{center}
     \includegraphics[clip,width=8.0cm]{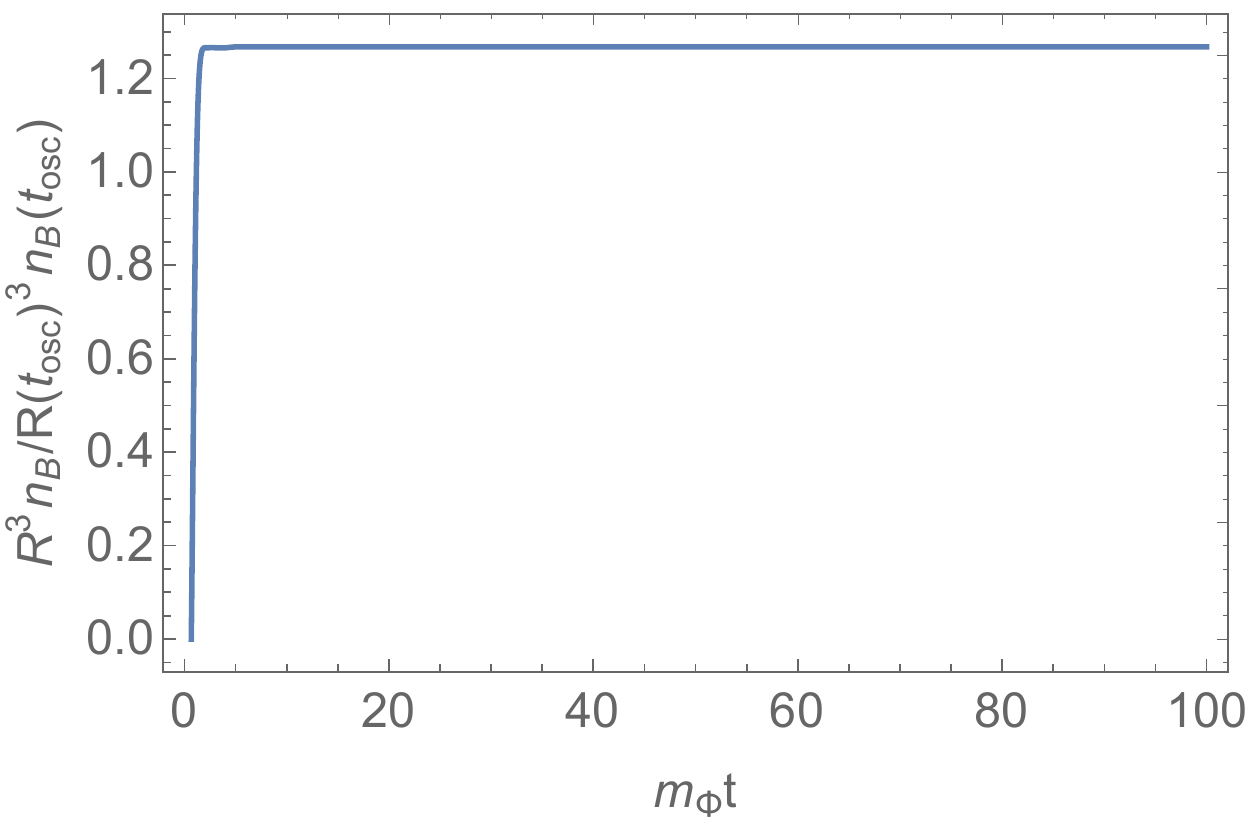}
    \end{center}
 \caption{The time evolution of the ratio between the numerical value $R(t)^3n(t)$~(\ref{eq-nint}) and $R^3(t_{\rm osc})n(t_{\rm osc})$~(\ref{LD}) for $c_1 = c_3=1$. The horizontal axis corresponds to $m_{\phi}t$. Here we set $\epsilon=2.4 \times 10^{-1}$ and $\delta_{\rm eff}=1$. The parameters are the same as in Fig.~\ref{fig:oscillate_2}. }
 \label{fig:BA_2}
 \end{figure}

Similarly, we numerically solve Eq.~(\ref{EOM_Hm}) and draw the trajectories of $S_1, \phi$ in Fig.~\ref{fig:oscillate_2} for the $c_1=c_3$ case. 
Here we set $c_1=c_3=1$, and the values of the other parameters are the same as in Fig.~\ref{fig:oscillate}. The initial conditions of $S_1$, $\phi$ are set by the values of $s_1,\chi$ at $z=18$. Then we can estimate $\epsilon \simeq 2.4 \times 10^{-1}$ numerically, and the ratio of the numerical value in Eq.~(\ref{eq-nint}) to the analytical one in Eq.~(\ref{LD}) is shown in Fig.~\ref{fig:BA_2}.  We also find that the numerical value  coincides with the analytical one in Eq.~(\ref{LD}) with $\delta_{\rm eff}=1$ after the AD/PQ fields start to oscillate. In this case, the mass of $\theta_\phi$ is comparable to the mass of $\theta_{S_1}$ because the field value of $\hat{\phi}$ is the same order as the one of $\hat{S}_1$. As a result, the phases of $S_1$ and $\phi$ rotate at the same time (drawn in Fig.~\ref{fig:oscillate_2}).

\clearpage

The ratio of the baryon number density $n_B$ to the entropy density $s$ 
after the reheating $(t=t_{\rm reh})$ is 
\begin{align}
\frac{n_B}{s}&=\frac{1}{s(t_{\rm reh})}\biggl(\frac{R(t_{\rm osc})}{R(t_{\rm reh})} \biggl)^3n_B(t_{\rm osc}) \nonumber \\
&=\frac{ \epsilon \ham m_{3/2}\delta_{\rm eff}T_{\rm reh}}{12M_P^2m_{\phi}^2}\biggl(\frac{m_{\phi}M_P^3}{\hat{\gamma}}\biggl)^{\frac{1}{2}},
\end{align}
where $T_{\rm reh}$ is the reheating temperature. Then the AD field decays and its energy density converts into  radiation~\cite{ADdecay}. 
Thus, the baryon asymmetry is estimated as
\begin{align}
\frac{n_B}{s}&=\frac{\epsilon \ham m_{3/2}\delta_{\rm eff}T_{\rm reh}}{12M_P^2m_{\phi}^2}\biggl(\frac{m_{\phi}M_P^3}{\hat{\gamma}}\biggl)^{\frac{1}{2}} \nonumber \\
&\simeq \left\{
\begin{array}{l}
0.53\times10^{-10}\biggl(\cfrac{T_{\rm reh}}{10^5~{\rm GeV}}\biggl)\biggl(\cfrac{1}{\hat{\gamma}}\biggl)^{\frac{1}{2}}\biggl(\cfrac{m_{3/2}}{m_{\phi}} \biggl)\biggl(\cfrac{m_{\phi}}{1~{\rm TeV}}\biggl)^{-\frac{1}{2}} \ \ \  {\rm for} \ \  c_1=\frac{1}{4},\ c_3=\frac{1}{5},\ \epsilon=3.1 \times 10^{-4} \\
0.40\times10^{-10}\biggl(\cfrac{T_{\rm reh}}{10^2~{\rm GeV}}\biggl)\biggl(\cfrac{1}{\hat{\gamma}}\biggl)^{\frac{1}{2}}\biggl(\cfrac{m_{3/2}}{m_{\phi}} \biggl)\biggl(\cfrac{m_{\phi}}{1~{\rm TeV}}\biggl)^{-\frac{1}{2}} \ \  {\rm for} \ \  c_1=c_3=1,\ \epsilon=2.4 \times 10^{-1}
\\
\end{array}
\right.
,
\label{BA}
\end{align}
where we set $\ham=1$ and $\delta_{\rm eff}\simeq 1$. 
It is remarkable that the obtained baryon asymmetry is very consistent with its current observed value and 
the tiny neutrino mass, simultaneously. 
The amount of baryon asymmetry is different from the AD mechanism without R-parity~\cite{ADwithR-parity} 
due to the nontrivial dynamics of PQ fields. 


Finally, we comment on the Q-ball problem~\cite{Qball1}. If the potential of the AD field $\phi$ is flatter than the quadratic one, 
the AD fields form the nontopological solitons called Q-balls~\cite{Qball2,Qball3}. In this paper, we assume a gravity-mediated SUSY-breaking scenario without R-parity. In gravity-mediated scenarios, one problem comes from the long lifetime of Q-balls. If R-parity is conserved in this scenario, Q-balls are unstable and the late time decays of Q-balls often overproduce the lightest supersymmetric particles (LSPs), which are candidates of cold dark matter. Since R-parity is violated in our model, the overabundance of the LSPs could be avoided due to the LSP decays. We leave the detailed calculation as a future work.

\subsection{Saxion decay in the SUSY DFSZ model}
In this section, we discuss the dynamics of the saxion $S_1$ in more detail. 
During inflation, the energy density of the saxion field should be smaller than that of the inflaton field, 
which constrains the field values of $S_1, \phi$ during inflation,
\begin{align}
\rho_{S_1}, \rho_\Phi \sim H_{\inf}^2 \phi^2 < \rho_{\rm inf}\simeq 3H_{\rm inf}^2 M_{P}^2.
\label{eq:constinf}
\end{align}
Now we have used the field values (\ref{eq-init}) during inflation and the masses of $S_1,\phi$ are 
of ${\cal O}(H_{\rm inf})$ as mentioned below Eq.~(\ref{eq:massesinf}). 
Thus, from Eqs.~(\ref{eq-init}) and (\ref{eq:constinf}), the field values of $S_1,\phi$ during inflation have to satisfy $S_1,\phi< M_{P}$.

After inflation, the inflaton and saxion fields oscillate around their minimum and their energy densities decay 
in proportion to $R^{-3}$. 
Consequently, the inflaton decays at the time
\begin{align}
&t_{\rm reh}\simeq \frac{1}{\Gamma_{\rm inf}} \simeq \left(\frac{90}{\pi^2 g_\ast}\right)^{1/2}\frac{M_{P}}{(T_{\rm reh})^2} \simeq 7\times 10^7\,{\rm GeV}^{-1} \left(\frac{106.75}{g_\ast}\right)^{1/2}\left(\frac{10^5\,{\rm GeV}}{T_{\rm reh}}\right)^2,
\end{align}
where $g_\ast$ denotes the effective degrees of freedom and $\Gamma_{\rm inf}$ is the total decay width of the inflaton. 
On the other hand, the total decay width of saxion field depends on the sparticle spectrum. (For more details, see e.g. Ref.~\cite{Bae:2013hma}.) 
When $m_{S_1}>2\mu$, the saxion decays mainly into Higgsino through the $\mu$-term in Eq.~(\ref{eq:WMSSM}). 
If such a decay is kinematically disallowed, the total decay width of saxion 
is dominated by the saxion decay into the CP-even Higgs bosons $h$, $H$ and the gauge bosons $W^\pm$ and $Z$. 
Note that saxion decays into axions are suppressed in our setup $\langle S_1\rangle \simeq \langle S_2\rangle \simeq f$, 
taking into account two PQ fields $S_{1,2}$ and singlet field $S_0$ with $U(1)_{\rm PQ}$ charges in Table~\ref{tab-PQcharge}~\cite{Chun:1995hc}.\footnote{For the dark radiation constraints in models 
with multiple PQ multiplets, we refer to e.g., Refs.~\cite{Bae:2013hma,Bae:2013qr}.} 
It then allows us to avoid the dark radiation problem from the saxion decay. 
As a result, the total decay width of the saxion\footnote{Now we consider the $\mu >m_A$ where $m_A$ is the mass of the CP-odd 
Higgs boson.}
\begin{align}
\label{eq:totdecS1}
\Gamma_{\rm tot}^{(S_1)}\simeq 
\left\{
\begin{array}{c}
\frac{1}{4\pi}\left(\frac{\mu_0}{f}\right)^2m_{S_1}\,\,\,\,(m_{S_1}>2\mu_0) \\
\frac{7}{2\pi}\frac{\mu_0^4}{f^2m_{S_1}}\,\,\,\,(m_{S_1}<2\mu_0) \\
\end{array}
\right.
\end{align}
determines the decay temperature of the saxion 
\begin{align}
T_{\rm dec}^{(S_1)} \simeq 
\left\{
\begin{array}{c}
4.1\times 10^2\ {\rm GeV} \left(\frac{106.75}{g_\ast}\right)^{1/4} \left(\frac{\mu_0}{10^4\ {\rm GeV}}\right)
\left(\frac{10^{12}\ {\rm GeV}}{f}\right)\left(\frac{m_{S_1}}{ 3\times 10^4\ {\rm GeV}}\right)^{1/2}\,\,\,\,(m_{S_1}>2\mu) \\
2.8\times 10^3\ {\rm GeV} \left(\frac{106.75}{g_\ast}\right)^{1/4} \left(\frac{\mu_0}{10^4\ {\rm GeV}}\right)^2
\left(\frac{10^{12}\ {\rm GeV}}{f}\right)\left(\frac{m_{S_1}}{ 10^3\ {\rm GeV}}\right)^{1/2}\,\,\,\,(m_{S_1}<2\mu) \\
\end{array}
\right.
,
\end{align}
and the decay time of the saxion
\begin{align}
t_{\rm dec}^{(S_1)} \simeq \frac{1}{\Gamma_{\rm tot}^{(S_1)}} \simeq \left(\frac{90}{\pi^2 g_\ast}\right)^{1/2} \frac{M_{P}}{(T_{\rm dec}^{(S_1)})^2} \simeq 7\times 10^{11}\,{\rm GeV}^{-1} \left(\frac{106.75}{g_\ast}\right)^{1/2}\left(\frac{10^3\,{\rm GeV}}{T_{\rm dec}^{(S_1)}}\right)^2.
\end{align}
Let us examine whether or not the saxion decay dilutes the baryon asymmetry via the entropy production from the 
saxion decay. The entropy dilution factor is determined by the ratio of the saxion decay temperature $T_{\rm dec}^{(S_1)}$ 
and the saxion-radiation equality temperature $T_{\rm eq}$, that is $T_{\rm eq}/T_{\rm dec}^{(S_1)}$. 
When the energy density of saxion is equal to that of radiation, 
$T_{\rm eq}$ is given by
\begin{align}
T_{\rm eq}\simeq \frac{1}{6}T_{\rm reh} \left(\frac{\hat{S}_1^0}{M_{P}}\right)^2
\simeq \frac{1}{6}T_{\rm reh} \left(\frac{H_{\rm inf}}{M_{P}}\right)^{1/2},
\end{align}
where $\hat{S}_1^0$ is the VEV of $\hat{S}_1$ during inflation.
In the setup discussed so far, the amplitude of saxion $\hat{S}_1^0\simeq (H_{\rm inf}M_{P}^3)^{1/4}$ is of ${\cal O}(10^{-2}\,M_{P})$ 
with $H_{\rm inf}=10^{11}\,{\rm GeV}$, 
and then $T_{\rm eq}$ is not larger than $T_{\rm dec}^{(S_1)}$ unless $T_{\rm reh}>{\cal O}(10^{15}\,{\rm GeV})$. 
Hence, there is no entropy dilution.

Finally, we comment on the lepton asymmetry generated from the saxion decay. 
The saxion field $S_1$ would be identified with the right-handed Majorana neutrino, as seen in the superpotential~(\ref{SRPV}). 
The leptogenesis scenario decaying from the Majorana neutrino has been discussed in the thermal~\cite{Fukugita:1986hr} and 
nonthermal epoch~\cite{Asaka:1999yd}. 
Since, in our case, the saxion oscillates around its minimum soon after inflation, 
it is possible to generate the lepton asymmetry through the coupling
\begin{align}
W \simeq 3\left(\frac{\langle S_1\rangle}{M_{P}}\right)^2 S_1 L H_u.
\label{eq:WS1LHu}
\end{align}
Such a lepton asymmetry is determined by
\begin{align}
\frac{n_L^{(S_1)}}{s}\simeq \delta \frac{n_{S_1}}{s},
\end{align}
where $\delta$ involves a CP asymmetry and lepton number violating factor determined by the saxion decay at one loop level. 
At the reheating era, the number density of saxion $n_{S_1}$ becomes
\begin{align}
\frac{n_{S_1}}{s}\simeq \frac{n_{S_1}(t_{\rm osc}^{(S_1)})}{s(t_{\rm reh})}\frac{\rho(t_{\rm reh})}{\rho (t_{\rm osc}^{(S_1)})} 
\simeq \frac{3}{4}\frac{T_{\rm reh}m_{S_1}(t_{\rm osc}^{(S_1)})}{H_{\rm inf}^2}\left(\frac{\hat{S}_1^0}{M_{\rm P}}\right)^2
\simeq 1.5\times 10^{-10}\left(\frac{T_{\rm reh}}{10^5\,{\rm GeV}}\right)\left(\frac{10^{11}\,{\rm GeV}}{H_{\rm inf}}\right)^{1/2},
\end{align}
where we use $m_{S_1}(t_{\rm osc}^{(S_1)})\simeq H_{\rm inf}$ and $\hat{S}_1^0\simeq 10^{-2}M_{P}$. 
Since the lepton asymmetry $n_L^{(S_1)}/s$ is further suppressed by the factor $\delta$ which is proportional to the 
effective Yukawa coupling $(\langle S_1\rangle/M_{P})^2$ in Eq.~(\ref{eq:WS1LHu}), 
the saxion produced lepton asymmetry is negligible even when $T_{\rm reh}>10^5\,{\rm GeV}$. 

In addition, we have to estimate the lepton asymmetry, taking into account the $S_1$ asymmetry which is generated by the AD mechanism as shown in Figs.~\ref{fig:noAterm}, \ref{fig:noAterm2}, \ref{fig:oscillate}, and \ref{fig:oscillate_2}. 
The $S_1$ asymmetry is numerically estimated as
\begin{align}
\frac{n_{S_1}-n_{\overline{S}_1}}{s}\simeq \frac{n_B}{s}\simeq 10^{-10},
\label{S1D}
\end{align}
where $n_{\overline{S}_1}$ is the number density of anti $S_1$. 
Then this $S_1$ asymmetry can be converted to lepton asymmetry due to the saxion decay at tree level. This asymmetry is determined by
\begin{align}
\frac{n^{(S_1)}_L}{s} \simeq \delta^{\prime}\frac{n_{S_1}-n_{\overline{S}_1}}{s}\simeq 10^{-10}\delta^{\prime},
\label{SLD}
\end{align}
where $\delta^{\prime}$ is a lepton number violating factor determined by the saxion decay at tree level. We obtain $\delta^{\prime}$ as 
\begin{align}
\delta^{\prime}=\frac{\Gamma^{(S_1)}_L}{\Gamma^{(S_1)}_{\rm tot}},
\label{SLD2}
\end{align}
where $\Gamma^{(S_1)}_L$ is the decay width of the saxion coming from the lepton number violating channel thorough the coupling (\ref{eq:WS1LHu}),
\begin{align}
\Gamma^{(S_1)}_L \simeq
\left\{
\begin{array}{c}
\sum_i\frac{1}{4\pi}\left(\frac{\mu_i}{f}\right)^2m_{S_1}\,\,\,\,(m_{S_1}>\mu_0) \\
\sum_i\frac{1}{8\pi}\frac{\mu_i^4}{f^2m_{S_1}}\,\,\,\,(m_{S_1}>m_{\tilde{L}}) \\
\end{array}.
\right.
\label{SLD3}
\end{align}
The first line denotes the saxion decays to Higgsinos and leptons, whereas the second line represents the saxion decays to 
Higgs and sleptons with mass $m_{\tilde{L}}$. 
For $f \ll M_P$ , the lepton number violating factor $\delta^{\prime}$ is estimated as
\begin{align}
\delta^{\prime}=\frac{\Gamma^{(S_1)}_L}{\Gamma^{(S_1)}_{\rm tot}} \lesssim \biggl(\frac{f}{M_P} \biggl)^2 \ll1.
\end{align}
As a result, this lepton asymmetry (\ref{SLD}) is also negligible.\footnote{If $S_1$ has a lepton number, $S_1$ asymmetry can be converted to lepton asymmetry. However, the dominant decay channel of the saxion $S_1$ is given by Eq.~(\ref{eq:totdecS1}) where the final states do not have the lepton number and it washes out lepton asymmetry originating from $S_1$ asymmetry. 
Thus, the lepton asymmetry through $S_1$ asymmetry is determined by Eqs. (\ref{SLD}), (\ref{SLD2}), (\ref{SLD3}).}


\subsection{Axion isocurvature perturbation}
\label{subsec:iso}
In this model, the massless Nambu-Goldstone boson called an axion exists through the spontaneous symmetry breaking of the $U(1)_{\rm PQ}$. This axion is a candidate for the dark matter and the present axion energy density is given by~\cite{DMabundance}
\begin{align}
\Omega_ah^2 \simeq 0.18\theta_a^2\biggl(\frac{f_a}{10^{12}~{\rm GeV}} \biggl)^{1.19},
\end{align}
where $\theta_a$ is the misalignment angle of the axion. $f_a$ is the axion decay constant 
depending on the domain wall number $N_{\rm DW}$, which is 6 for the DFSZ axion model~\cite{DFSZaxion},
\begin{align}
f_a=\frac{\sqrt{2}f}{N_{\rm DW}}.
\end{align}
For $f=10^{12}\ {\rm GeV}$ and $\theta_a=1.9$, the axion energy density is coincident with the dark matter energy density $\Omega_{\rm CDM} \simeq 0.12$.

However, such a massless boson would have been problematic in the early Universe\cite{AxionProblem}. 
In our model, the $U(1)_{\rm PQ}$ symmetry is spontaneously broken during inflation and is not recovered after inflation. 
Because of $U(1)_{\rm PQ}$ symmetry breaking, the domain wall problem~{\cite{domainwall1,domainwall2} does not occur. The PQ field $S_1$ gets the large VEV $\langle S_1 \rangle \simeq \langle \phi \rangle \simeq (H_{\rm inf}M_P^3)^{1/4}$ during inflation. It can suppress the axion isocurvature perturbation~\cite{Aiso1,Aiso2,Aiso3,Aiso4}. The axion almost consists of the linear combination 
of $\theta_{S_1}$ and $\theta_\phi$ for $|S_1|,|\phi| \gg f$ during inflation, so the axion $a$ takes the form,
\begin{align}
a \simeq \frac{1}{\sqrt{\langle \hat{S_1} \rangle^2+\langle \hat{\phi} \rangle^2}}(\langle \hat{S_1} \rangle^2 \theta_{S_1} - \langle \hat{\phi} \rangle^2 \theta_{\phi}).
\end{align}
The PQ breaking scale $v$ during inflation is
\begin{align}
v\simeq \max [\langle \hat{S}_1 \rangle, \langle \hat{\phi} \rangle ] \simeq (H_{\rm inf}M_P^3)^{1/4}.
\end{align}

The power spectrum of cold dark matter (CDM) isocurvature perturbation $\mathcal{P}_{\rm iso}$ is
\begin{align}
\mathcal{P}_{\rm iso}\simeq r^2\biggl(\frac{H_{\rm inf}}{\pi v\theta_a} \biggl)^2,
\end{align}
where $r$ is the ratio of the present axion energy density to the matter energy density, $r=\Omega_ah^2/\Omega_mh^2$. 
The Planck constraint on the uncorrelated isocurvature perturbation~\cite{Planck2018} becomes
\begin{align}
H_{\rm inf} \lesssim 2.2 \times10^{12}~{\rm GeV}\theta_a^{-1}\biggl(\frac{10^{12}~{\rm GeV}}{f_a} \biggl)^{1.59}.
\end{align}
Thus, the axion isocurvature perturbation is mildly suppressed due to the large $v$. In our analysis, we consider $H_{\rm inf}\simeq 10^{11}~{\rm GeV}$ to avoid this constraint.

\section{Conclusion and discussion}
\label{sec:con}
In this paper, we have investigated the baryon asymmetry in the SUSY DFSZ axion model without R-parity. 
Such R-parity violating interactions are motivated not only by explaining  the tiny neutrino masses, but also by 
avoiding the cosmological gravitino and moduli problems. 
In this model, the Affleck-Dine mechanism can work out via the coupling among PQ fields and sleptons (squarks). 
We reveal that the R-parity violating terms produce the appropriate amount of baryon asymmetry in the parameter region, 
explaining the axion dark matter abundance, the smallness of $\mu$- and $R$-parity violating interactions, and the atmospheric mass-squared difference of neutrinos. 
Furthermore, in this model, the constraint for the Hubble parameter during inflation is relaxed because the PQ breaking scale is enhanced during inflation. 

Although, in this paper, we have focused on the atmospheric mass-squared difference of neutrinos, it would be interesting to discuss 
in more detail the neutrino masses and flavor mixings which will be the subject of future work. 
The SUSY DFSZ axion model without R-parity may explain the structure of the neutrino masses and flavor mixings without severe tunings of parameters.


\section*{Acknowledgments}
We thank J. Kawamura for the useful comments and for improving the draft of this paper. We also thank T. Moroi for the useful comments, and the referees for suggesting improvements and pointing out errors. 
K.~A. is grateful to M. Yamaguchi for the valuable discussion and also thanks K. Fujikura and M. Ibe for the useful comments.
H.~O. was supported in part by a Grant-in-Aid for Young Scientists (B) (No.~17K14303) from the Japan Society for the Promotion of Science.

\appendix

\section{General coupling between AD/PQ fields and the minimum during inflation}
\label{app}

In Sec.3, we consider the PQ charges as Table~\ref{tab-PQcharge} and $\overline{u}\overline{d}\overline{d}$ D-flat direction. However, we can consider other PQ charges and other D-flat directions in the AD mechanism. Therefore, in this section, we consider the following general superpotential of the AD field and PQ fields $W_{\rm AD}(S_1,\Phi)$:
\begin{align}
W_{\rm AD}(S_1,\Phi)&=-\frac{\gamma^\prime S_1^m\phi^n}{nM_P^{n+m-3}},
\end{align} 
where $m,n$ are integers and $\gamma^\prime$ is a dimensionless parameter, giving rise to the following superpotential:
\begin{align}
W&=W_{\rm inf}(I)+W_{\rm AD}(S_1,\Phi)+W_{\rm PQ}+W'', \nonumber \\
W_{\rm PQ}&= \kappa S_0(S_1S_2-f^2), \nonumber \\
W''&=\biggl( \alpha''W_{\rm AD}(S_1,\phi)+\beta''W_{\rm PQ} \biggl)\frac{F_I^{\ast}}{M_P}+{\rm h.c.},
\end{align}
where $\alpha'', \beta''$ are coupling constants. Let us investigate the minimum during inflation in this setup. 
After inflation, we must numerically study the dynamics of the AD/PQ fields, and 
the results depend on the detail of parameters. 
In this respect, we will postpone the investigation of their dynamics for a future work, but 
the results will be similar to Secs.~\ref{subsec:3_1_3} and \ref{subsec:3_1_4}. 
Here we consider $n,m\geq 2$.
In our model presented in Sec.~3, namely $n=3,m=3$, the following calculation is simplified.
We can apply the following calculation to the R-parity conserving case. 

First, we consider the scalar potential for the AD/PQ fields.  We assume that the K$\ddot{{\rm a}}$hler potential is the same as in Eq.~(\ref{KP}). Then the scalar potential for the AD/PQ field is described as
\begin{align}
V&=V_{\rm Hubble}+V_{\rm soft}+V_{\rm A}'+V_{\rm F}', \nonumber \\
V_{\rm A}'&= (a_H'H+a_mm_{3/2})\frac{\gamma^\prime S_1^m\phi^n}{nM_P^{n+m-3}}+(b_H'H+b_m'm_{3/2})\kappa S_0(S_1S_2-f^2)+{\rm h.c.}, \nonumber \\
V_{\rm F}'&=|\kappa|^2|S_1S_2-f^2|^2+\biggl|\kappa S_0S_2-\frac{m\gamma^\prime S_1^{m-1}\phi^n}{nM_P^{n+m-3}} \biggl|^2+|\kappa|^2|S_0S_1|^2+\frac{|\gamma^\prime|^2|S_1|^{2m}|\phi|^{2(n-1)}}{M_P^{2n+2m-6}},
\end{align}
where $V_{\rm Hubble}$ and $V_{\rm soft}$ are the same as in Eqs.~(\ref{eq-Vhub}) and (\ref{eq-Vsoft}).

Let us investigate a minimum of the potential. Ignoring the soft supersymmetry-breaking effect, the phase-dependent term in this potential is 
 \begin{align}
 V_{\rm phase} &= -2\hat{\kappa}^2f^2\hat{S_1}\hat{S_2}\cos(\theta_{S_1}+\theta_{S_2})-2\hat{\kappa}\hat{\gamma}^\prime\hat{S_0}\hat{S_2}\frac{m\hat{S_1}^{m-1}\hat{\phi}^n}{nM_P^{n+m-3}}\cos(\theta_{S_0}+\theta_{S_2}-(m-1)\theta_{S_1}-n\theta_{\phi}+\zeta') \nonumber \\
&\ \ \ \ -2\hat{\kappa}\hat{b_H'}(H\hat{S_0})\hat{S_1}\hat{S_2}\cos(\theta_{S_0}+\theta_{S_1}+\theta_{S_2}+\eta')+2\hat{\kappa}\hat{b_H'}f^2(H\hat{S_0})\cos(\theta_{S_0}+\eta') \nonumber \\
&\ \ \ \ -2\hat{\gamma}^\prime\hat{a}_HH\frac{\hat{S_1}^m\hat{\phi}^n}{nM_P^{n+m-3}}\cos(m\theta_{S_1}+n\theta_{\phi}+\xi'),
 \end{align}
 where $\zeta', \eta', \xi'$ are some numerical constants. Then the minimum of the phases are
 \begin{align}
 \langle\theta_{S_1}+\theta_{S_2}\rangle&\simeq0, \nonumber \\
\langle\theta_{S_0}+\theta_{S_2}-(m-1)\theta_{S_1}-n\theta_{\phi}\rangle&\simeq -\zeta', \nonumber \\
\langle m\theta_{S_1}+n\theta_{\phi}\rangle&\simeq -\xi'.
 \end{align}
 We assume that $H|S_0| < f^2$ which is confirmed in the same way as in Sec. 3.1.2. We set all dimensionless parameters as $\mathcal{O}(1)$. Ignoring the soft mass terms, the first derivatives in each radial direction of field at the minima of the phase directions are
 \begin{align}
 \frac{\partial V}{\partial \hat{S_0}}&\simeq2\hat{\kappa}\hat{S_2}\biggl(\hat{\kappa}\hat{S_0}\hat{S_2}-\frac{m\hat{\gamma}^\prime\hat{S_1}^{m-1}\hat{\phi}^n}{nM_P^{n+m-3}}\biggl)+2\hat{\kappa}^2\hat{S_0}\hat{S_1}^2+2c_0H^2\hat{S_0}, \nonumber \\
\frac{\partial V}{\partial \hat{S_1}}&\simeq2\hat{\kappa}^2\hat{S_2}\biggl(\hat{S_1}\hat{S_2}-f^2\biggl)-\frac{2m(m-1)\hat{\gamma}^\prime\hat{S_1}^{m-2}\hat{\phi}^n}{nM_P^{n+m-3}}\biggl(\hat{\kappa}\hat{S_0}\hat{S_2}-\frac{m\hat{\gamma}^\prime\hat{S_1}^{m-1}\hat{\phi}^n}{nM_P^{n+m-3}} \biggl) \nonumber \\
&\ \ \ \ +2\hat{\kappa}^2\hat{S_0}^2\hat{S_1}+\frac{2m(\hat{\gamma}^\prime)^2\hat{S_1}^{2m-1}\hat{\phi}^{2(n-1)}}{M_P^{2n+2m-6}}-\frac{2m\hat{\gamma}^\prime \hat{a}_HH\hat{S_1}^{m-1}\hat{\phi}^n}{nM_P^{n+m-3}}-2c_1H^2\hat{S_1}, \nonumber \\
\frac{\partial V}{\partial \hat{S_2}}&\simeq2\hat{\kappa}^2\hat{S_1}\biggl(\hat{S_1}\hat{S_2}-f^2\biggl)+2\hat{\kappa}\hat{S_0}\biggl( \hat{\kappa}\hat{S_0}\hat{S_2}-\frac{m\hat{\gamma}^\prime\hat{S_1}^{m-1}\hat{\phi}^n}{nM_P^{n+m-3}} \biggl)+2c_2H^2\hat{S_2}, \nonumber \\
\frac{\partial V}{\partial \hat{\phi}}&\simeq-\frac{2m\hat{\gamma}^\prime\hat{S_1}^{m-1}\hat{\phi}^{n-1}}{M_P^{n+m-3}}\biggl(\hat{\kappa}\hat{S_0}\hat{S_2}-\frac{m\hat{\gamma}^\prime\hat{S_1}^{m-1}\hat{\phi}^n}{nM_P^{n+m-3}} \biggl)+\frac{2(n-1)(\hat{\gamma}^\prime)^2\hat{S_1}^{2m}\hat{\phi}^{2n-3}}{M_P^{2n+2m-6}} \nonumber \\
&\ \ \ \ -\frac{2\hat{\gamma}^\prime \hat{a}_HH\hat{S_1}^m\hat{\phi}^{n-1}}{M_P^{n+m-3}}-2c_3H^2\hat{\phi}.
 \end{align}
From the extremal conditions $\frac{\partial V}{\partial \hat{\phi_i}}=0$, one of the extrema is given by
\begin{align}
\langle\hat{S_0}\rangle&=\frac{\hat{\kappa} \langle\hat{S_2}\rangle}{\hat{\kappa}^2\langle\hat{S_1}\rangle^2+\hat{\kappa}^2 \langle \hat{S_2}\rangle^2+c_0H^2}\frac{m\hat{\gamma}^\prime \langle \hat{S_1}\rangle^{m-1} \langle\hat{\phi} \rangle^n}{nM_P^{n+m-3}}, \nonumber \\
\langle \hat{S_2} \rangle&\simeq\frac{f^2}{\langle \hat{S_1} \rangle}, \ \ \ \
\langle\hat{S_1}\rangle\simeq k'\langle \hat{\phi} \rangle, \nonumber \\
\langle \hat{\phi} \rangle&\simeq\biggl(\frac{k'^{m-2}\frac{m}{n}\hat{a}_H+\sqrt{k'^{2m}\frac{m^2}{n^2}\hat{a}_H^2+4c_1(k'^{2m-4}\frac{m^2(m-1)}{n^2}+k'^{2m-2}m)}}{2(k'^{2m-4}\frac{m^2(m-1)}{n^2}+k'^{2m-2}m)}\frac{HM_P^{n+m-3}}{\hat{\gamma}^\prime} \biggl)^{\frac{1}{n+m-2}},
\end{align}
where $k'$ is some numerical constant which depends on $\hat{a}_H, c_1, c_2, n$, and $m$. 
To get this extrema, we must satisfy the following condition during inflation,
\begin{align}
\langle \hat{S_1} \rangle \gtrsim f.
\label{S1condition}
\end{align}
Note that in Sec.3, Eq.~(\ref{S1condition}) is satisfied under $H>m_{\phi}$ and $f \simeq 10^{12}~{\rm GeV}$. From the potential and Eq.(\ref{S1condition}), $S_0$ and $S_2$ obtain large positive masses of $\langle \hat{S_1} \rangle$. Thus, we assume that $\hat{S_0}, \hat{S_2}$, and all phase directions are fixed at the minimum. Then the mass matrix of $\hat{S_1}$ and $\hat{\phi}$ is
\begin{align}
\frac{1}{2}
    \begin{pmatrix}
   \dfrac{\partial^2 V}{\partial \hSo \partial \hSo} & 
   \dfrac{\partial^2 V}{\partial \hSo \partial \hph } \\[2.5ex]
   \dfrac{\partial^2 V}{\partial \hph \partial \hSo} & 
   \dfrac{\partial^2 V}{\partial \hph \partial \hph}
    \end{pmatrix},
\label{eq:massmatrix_general} 
\end{align}
where
\begin{align}
\frac{1}{2}\frac{\partial^2 V}{\partial\hat{S_1} \partial\hat{S_1}} &\simeq \frac{m^2(m-1)(2m-3)(\hat{\gamma}^\prime)^2\hat{S_1}^{2m-4}\hat{\phi}^{2n}}{n^2M_P^{2n+2m-6}}+\frac{m(2m-1)(\hat{\gamma}^\prime)^2\hat{S_1}^{2(m-1)}\hat{\phi}^{2(n-1)}}{M_P^{2n+2m-6}} \nonumber \\
&\ \ \ \ -\frac{m(m-1)\hat{\gamma}^\prime \hat{a}_HH\hat{S_1}^{m-2}\hat{\phi}^n}{nM_P^{n+m-3}}-c_1H^2,  \label{mass1} \\
\frac{1}{2}\frac{\partial^2 V}{\partial\hat{S_1}\partial\hat{\phi}} &\simeq \frac{2m^2(m-1)(\hat{\gamma}^\prime)^2\hat{S_1}^{2m-3}\hat{\phi}^{2n-1}}{nM_P^{2n+2m-6}}+\frac{2(n-1)m(\hat{\gamma}^\prime)^2\hat{S_1}^{2m-1}\hat{\phi}^{2n-3}}{M_P^{2n+2m-6}} \nonumber \\
&\ \ \ \ -\frac{m\hat{\gamma}^\prime \hat{a}_HH\hat{S_1}^{m-1}\hat{\phi}^{n-1}}{M_P^{n+m-3}}, \label{mass2} \\
\frac{1}{2}\frac{\partial^2 V}{\partial\hat{\phi} \partial\hat{\phi}} &\simeq \frac{m^2(2n-1)(\hat{\gamma}^\prime)^2\hat{S_1}^{2(m-1)}\hat{\phi}^{2(n-1)}}{nM_P^{2n+2m-6}}+\frac{(2n-3)(n-1)(\hat{\gamma}^\prime)^2\hat{S_1}^{2m}\hat{\phi}^{2n-4}}{M_P^{2n+2m-6}} \nonumber \\
&\ \ \ \ -\frac{(n-1)\hat{\gamma}^\prime \hat{a}_HH\hat{S_1}^m\hat{\phi}^{n-2}}{M_P^{n+m-3}}-c_3H^2. 
\label{mass3}
\end{align} 
One can realize the positive eigenvalues of Eq.~(\ref{eq:massmatrix_general}) under $|a_H| \gg c_1,c_3$.



\end{document}